\documentclass[a4paper,11pt,english]{article}

\usepackage{amsfonts,amssymb,amsthm}    
\usepackage[affil-it,auth-sc]{authblk}          
            
\usepackage{bigints}
\usepackage{bm}                         
\usepackage{braket}              
\usepackage[utf8]{inputenc}
\usepackage{caption}
\usepackage{enumerate}
\usepackage{enumitem}           
\usepackage[left=27mm,right=27mm,top=27mm,bottom=29mm]{geometry}           
\usepackage{latexsym}            
\usepackage{mathrsfs}           
\usepackage{mathtools}                  
\usepackage{mparhack,fixltx2e,relsize}     
\usepackage[only,llbracket,rrbracket]{stmaryrd} 
\usepackage{subfig}                      
\usepackage{cite}
\usepackage{upgreek}
\usepackage[english]{varioref}   
\usepackage[english]{babel}                   
\usepackage{graphicx}
\usepackage{color}
\usepackage{caption}
\usepackage[titletoc,toc,title]{appendix}
\usepackage{hyperref}
\usepackage{bookmark}
\usepackage{lmodern}
\usepackage{multirow}
\usepackage{float} 

\hypersetup{
  breaklinks=true,
  bookmarksnumbered,
  bookmarksopen=true, 
  bookmarksopenlevel=2, 
  pdftitle={Scattering of Antiplane Shear Waves by Fractals in Strain Gradient Elasticity},
  pdfauthor={E. Alevras, Th. Zisis, P.A. Gourgiotis},
  pdfkeywords={Koch snowflake, microstructure, metamaterials, wave trapping, Green's function} }

\begin{document}
\selectlanguage{english}

\title{Scattering of Antiplane Shear Waves by Fractals in Strain Gradient Elasticity}

\author[1]{E. Alevras}
\author[1]{Th. Zisis}
\author[1]{P.A. Gourgiotis\footnote{Corresponding author:\,e-mail:\,pgourgiotis@mail.ntua.gr.}}
\affil[1]{ Mechanics Division, School of Applied Mathematical and Physical Sciences, National Technical University of Athens, Zographou, 15773, Greece}

\date{}
\maketitle

\begin{abstract}
\noindent
Wave manipulation is essential in various applications, including seismic wave protection, sound isolation, and acoustic device design. This study examines the scattering and trapping of antiplane SH waves in a microstructured solid embedded with rigid pins. The material's response is governed by the theory of strain gradient elasticity. A method is proposed for identifying the wavenumbers that induce resonance in the elastic body, based on specific material parameters and pin configurations. The analysis focuses on a system featuring a Koch snowflake–type pin layout, a fractal curve generated through an iterative process. This geometry allows the exploration of a complex arrangement characterized by a high concentration of sharp corners that promote scattering. The system's response to this self-similar configuration is analysed and compared to circular pin arrangements with an equivalent number of pins. A distinct resonance mode is identified, where the motion is trapped within the pin configuration and the conditions for its emergence are analyzed in detail. Furthermore, the study explores the influence of the characteristic lengths of the problem that are induced by the dynamic gradient elasticity theory. The findings indicate that in fractal pin arrangements, the system's response is rather dictated by the fractal dimension rather than by the total number of pins, highlighting the significance of geometry in wave propagation dynamics.
\end{abstract}


\section{Introduction}
\noindent
This study investigates the propagation of SH antiplane elastic waves in an infinite isotropic solid, perturbed by rigid pins arranged along a fractal curve. The material's response is governed by Toupin–Mindlin strain gradient elasticity which accounts for microstructural effects influencing the wave behavior. The fractal geometry considered in detail is the Koch snowflake. This particular fractal is chosen due to its highly intricate boundary, characterized by numerous sharp corners and cusps. These geometric features enhance wave trapping by promoting multiple internal reflections within confined regions of the structure, thereby inducing strong localized responses. Moreover, using a fractal geometry rather than an arbitrary complex shape enables a systematic comparison of the system's response across successive iterations. This allows us to examine whether the wave behavior converges toward a specific pattern as the geometric complexity increases. We specifically analyse how the geometry and fractal dimension of the configuration influence the wave propagation, presenting results up to the fourth iteration of the Koch snowflake.

When travelling waves interact with fractal boundaries, they exhibit unusual behavior in terms of reflection, dispersion, and diffraction. These phenomena are collectively referred to in the literature as "diffractals", a term first coined by Berry \cite{berry1979diffractals} to describe the unique diffraction patterns and wave interactions that arise when waves encounter self-similar, non-smooth geometries. The complex interplay between the wave and the fractal structure results in intricate scattering dynamics and spectral features not present in smooth or regular geometries. Although diffractals have been extensively studied in the context of electrodynamics \cite{jaggard1990fractal, jaggard1991fractal, aguili2009study, bhattacharyya2016wave}, they have not yet been investigated in continuous mechanical systems.

Prior studies on antiplane shear wave propagation have significantly advanced our understanding of scattering and localization phenomena in structured media, particularly in periodic and quasi-periodic systems \cite{andrianov2021linear, movchan2017mathematical}. Analytical techniques for multiple scattering \cite{parnell2011multiple}, high-frequency homogenization \cite{craster2010high, guenneau2013homogenization, touboul2024high}, and dispersion \cite{eremeyev2019comparison, eremeyev2020strongly} in elastic continua have been extensively developed, including systems containing discrete microstructures such as rigid pins or inclusions \cite{shodja2012surface, shodja2015scattering, delfani2024strain, lazaro2025weak}. Nonetheless, most of these studies have concentrated on regular geometries and have predominantly been carried out within the frameworks of classical elasticity or plate theory.

At this point, it is important to reference the work of Evans and Porter \cite{evans2007penetration} and Haslinger et al. \cite{haslinger2012transmission, haslinger2014localisation, haslinger2018localization}, who investigated elastic wave propagation in infinite Kirchhoff plates containing a finite number of pins arranged in relatively simple configurations. Their use of the Kirchhoff plate model enabled the imposition of localized constraints through point supports, a feature that is mathematically compatible with the structure of the theory. As Mindlin originally noted \cite{mindlin1964micro} (see also \cite{gavardinas2018karman}), there are formal analogies between Kirchhoff plate theory and certain formulations of strain gradient elasticity. However, the key distinction lies in the nature of the models: Kirchhoff theory is inherently structural and limited to thin-plate geometries, whereas strain gradient elasticity is a three-dimensional continuum theory. In this work, we aim to show that such localized effects can be naturally captured within a generalized continuum framework such as the Toupin-Mindlin theory of gradient elasticity.

In our case, the medium is subjected to an antiplane shear motion, implying that the propagating waves are shear horizontally polarized (SH) waves. The pins are modelled as points with null out-of-plane displacement. To construct the solution to the scattering problem, we employ a superposition approach based on the time-harmonic Green's function in strain gradient elasticity. The Green's function is derived in the present study in closed form  and represents the response of an infinite microstructured medium to an antiplane concentrated body force, serving as the fundamental building block of our methodology. A crucial feature that distinguishes this setting from its classical counterpart is that, in strain gradient elasticity, the antiplane Green's function yields finite displacements at the source point. The regularizing nature of the theory is essential to accommodate the superposition of multiple sources without introducing singularities. Moreover, the use of strain gradient elasticity enables a natural comparison between the physical scales of the problem such as the wavelengths of the propagating waves, the number and spacing of the pins and the intrinsic characteristic lengths of the theory, which are associated with micro-inertia and the strain energy. This scale interplay is crucial for capturing the influence of microstructure on wave propagation and scattering phenomena. Our findings show that the fractal geometry, combined with the microstructural scale effects inherent in strain gradient elasticity, leads to intricate wave-structure interactions that reveal important insights into wave trapping and localization. These results have promising implications for applications such as energy harvesting \cite{alshaqaq2020graded, de2020graded, de2021graded}, where the enhanced strain gradients generated by trapped waves inside the pin configurations may be exploited in the design of flexoelectric metamaterials \cite{giannakopoulos2023hyperbolicityI, knisovitis2024anti}. Further potential applications include vibration isolation \cite{badreddine2012broadband, miniaci2016large, achaoui2017clamped, brule2014experiments, brule2017flat}, and the design of acoustic metamaterials with tailored properties for controlled wave propagation and localization \cite{craster2012acoustic, nobili2020new,  movchan2022a, RIZZI2024105269, Madeo2024}.

The structure of the paper is as follows: \hyperref[sec2]{Section~2} presents the fundamental principles of Form II of Mindlin’s general theory of strain gradient elasticty, focusing on the derivation of the displacement equations of motion and the associated dispersion relations. \hyperref[sec3]{Section~3} derives the closed-form time-harmonic Green's function for an infinite body under antiplane shear motions, using Fourier transform analysis. In \hyperref[sec4]{Section~4}, we formulate the scattering problem where wave scattering emerges from a cluster of rigid pins and establish the conditions for resonance trapping modes that confine motion within the pinned region. \hyperref[sec5]{Section~5} provides a detailed analysis of a pinned configuration based on the Koch snowflake fractal. Among the examined parameters, the dynamic response of the system is shown to be primarily governed by the ratio of the two microstructural length scales introduced by the generalized continuum theory, which dictate dispersion behavior and the fractal dimension of the configuration, rather than the total number of pins. To highlight the effect of geometric complexity, comparative results are also presented for a circular pin arrangement containing the same number of pins as each iteration of the Koch snowflake.

\section{Fundamentals of form II of Mindlin's general theory}
\label{sec2}
\noindent
In this section, we briefly present the fundamental equations of Form II of Mindlin’s general theory, also known as strain gradient elasticity, with a focus on deriving the displacement equations of motion and the corresponding dispersion relations. Notably, these equations are common to all three formulations of Mindlin’s theory, and Form II is adopted here as a representative framework to illustrate the derivation process \cite{mindlin1964micro}.

In form II of Mindlin's general theory, the strain-energy density $W$ of an isotropic centrosymmetric elastic body is a quadratic form of the strains and the strain gradients,
 \begin{align}\label{eq.1}
W &= \frac{1}{2} \lambda \varepsilon_{ii} \varepsilon_{jj} + \mu \varepsilon_{ij} \varepsilon_{ij} + \hat{a}_1 \kappa_{iik} \kappa_{kjj} + \hat{a}_2 \kappa_{ijj} \kappa_{ikk} + \hat{a}_3 \kappa_{iik} \kappa_{jjk} +  \hat{a}_4 \kappa_{ijk} \kappa_{ijk} \nonumber \\
&+ \hat{a}_5 \kappa_{ijk} \kappa_{kji},
\end{align}
where $\varepsilon_{ij} = (1/2) (\partial_i u_j + \partial_j u_i)$ is the infinitesimal strain tensor, $\kappa_{ijk} = \partial_i \varepsilon_{jk} = \partial_i \varepsilon_{kj}$ is the strain gradient, $u_p$ is the displacement vector, $\lambda$, $\mu$ are the Lamé constants, $\hat{a}_1 - \hat{a}_5$ are additional material constants introduced by the strain gradient formulation to capture microstructural effects and are explicitly defined by Mindlin in \cite{mindlin1964micro}, and $\partial_p () = \partial()/\partial x_p$ with the Latin indices spanning the range $\left(1, 2, 3\right)$ (indicial notation and summation convention is used throughout).

The stresses are related to the deformation measures defined above through the following hyperelastic constitutive laws
\begin{equation}\label{eq.2}
\tau_{ij} \equiv \frac{\partial W}{\partial \varepsilon_{ij}}, \qquad \mu_{ijk} \equiv \frac{\partial W}{\partial \kappa_{ijk}} \equiv \frac{\partial W}{\partial \left( \partial_i \varepsilon_{jk} \right)} = \mu_{ikj} \, .
\end{equation}
Eq. \eqref{eq.1} along with the definitions \eqref{eq.2}, lead to the following constitutive relations:
\begin{equation} \label{eq.3}
\tau_{ij} = 2 \mu \varepsilon_{ij} + \lambda \varepsilon_{kk} \delta_{ij},
\end{equation}
\begin{align}
\label{eq.4}
\mu_{ijk} &= \frac{1}{2} \hat{a}_1 \left( \kappa_{kll} \delta_{ij} + 2 \kappa_{lli} \delta_{jk} + \kappa_{jll} \delta_{ki} \right) + 2 \hat{a}_2 \kappa_{ill} \delta_{jk} + \hat{a}_3 \left( \kappa_{llk} \delta_{ij} + \kappa_{llj} \delta_{ik} \right) \nonumber \\
&+ 2 \hat{a}_4 \kappa_{ijk} + \hat{a}_5 \left( \kappa_{kij} + \kappa_{jki} \right),
\end{align}
where $\tau_{ij}$ is the Cauchy (in Mindlin's notation) monopolar stress tensor, $\mu_{ijk}$ is the dipolar (double) stress tensor and, $\delta_{ij}$ is the Kronecker delta.

Following Toupin \cite{toupin64}, the kinetic energy density $T$ is defined as a function of the velocities and the gradients of the velocities
\begin{equation}\label{eq.5}
T = \frac{1}{2} \rho \dot{u}_i \dot{u}_i + \frac{1}{2} \rho h^2 (\partial_i \dot{u}_j) (\partial_i \dot{u}_j),
\end{equation}
where $\rho$ is the mass density of the material, the superposed dot denotes time derivative, and $h$ is a microstructural length associated with the micro-inertia of the medium. According to Mindlin \cite{mindlin1964micro}, $h^2=d^2/3$ where $d$ is the side length of the representative cell of the microstructure.

Further, employing Hamilton's principle and using Eqs. \eqref{eq.1} and \eqref{eq.5} the equations of motion and the boundary conditions are obtained as
\begin{gather}
\label{eq.6}
\partial_j \left( \tau_{jk} - \partial_i \mu_{ijk} \right) + F_k = \rho \ddot{u}_k - \rho h^2 \partial_{jj} \ddot{u}_k, \\
\label{eq.7}
P_k^{(n)} = n_j (\tau_{jk} - \partial_i \mu_{ijk}) - D_j ( n_i \mu_{ijk} ) + ( D_p n_p ) n_i n_j \mu_{ijk} + \rho h^2 n_j \partial_j \ddot{u}_k, \\
\label{eq.8}
R_k^{(n)} = n_i n_j \mu_{ijk}, \\
\label{eq.9}
E_k = \llbracket n_i m_j \mu_{ijk} \rrbracket ,
\end{gather}
where $F_k$ is the body force vector, $P_k^{(n)}$ is the monopolar traction vector, $R_k^{(n)}$ is the dipolar traction vector, $E_k$ is the edge force, $n_i$ is the outward unit vector vector, normal to the boundary $S$, $D \equiv n_k \partial_k$ is the normal gradient operator, and $D_i \equiv \partial_i - n_i D$ is the surface gradient operator. The non-classical boundary condition \eqref{eq.9} arises only when non-smooth boundaries are considered. The double brackets $\llbracket \rrbracket$ indicate the jump across a corner of the boundary $S$, which is the difference in the values of the enclosed quantity on either side of the edge, while $m_i$ is a tangential vector along the edge line.

The mechanical power per unit area expended on the boundary is expressed as  
\begin{equation}\label{power}
\mathcal{P} = P_k^{(n)} \dot{u}_k + R_k^{(n)} D \dot{u}_k \, ,
\end{equation}  
where the first term corresponds to the power due to monopolar tractions, and the second term accounts for the power expended by the dipolar tractions through the normal gradient of the velocity.

Combining the above equations, the displacement equations of motion are finally obtained in the form
\begin{equation}\label{eq.10}
\left( \lambda + 2 \mu \right) \left( 1 - \ell^2_1 \nabla^2 \right) \nabla \nabla \cdot \mathbf{u} - \mu \left( 1 - \ell^2_2 \nabla^2 \right) \nabla \times \nabla \times \mathbf{u} + \mathbf{F} = \rho \left( \mathbf{\ddot{u}} - h^2 \nabla^2 \mathbf{\ddot{u}} \right),
\end{equation}
where $\nabla^2$ is the three-dimensional Laplace operator. The intrinsic parameters $\ell^2_i$, $h^2$, which appear in Eq. \eqref{eq.10} have dimensions of $[length]^2$ and are defined as 
\begin{equation}\label{eq.11}
\ell^2_1 = \frac{2 \left( \hat{a}_1 + \hat{a}_2 + \hat{a}_3 + \hat{a}_4 + \hat{a}_5 \right)}{\lambda + 2 \mu}, \qquad \ell^2_2 = \frac{ \left( \hat{a}_3 + 2\hat{a}_4+\hat{a}_5 \right)}{2\mu}.
\end{equation}
In the limit $\left( \ell^2_i, h^2 \right) \rightarrow 0 $, the Navier-Cauchy equations of classical linear isotropic elasticity are recovered from Eq. \eqref{eq.10}. 

Next, by taking the divergence and curl of the homogeneous Eq. \eqref{eq.10} we obtain the equations governing the propagation of dilatation and rotation, respectively:
\begin{equation}\label{eq.12}
V^2_P \left( 1 - \ell^2_1 \nabla^2 \right) \nabla^2 \left( \nabla \cdot \mathbf{u} \right) = \left( 1 - h^2 \nabla^2 \right) \nabla \cdot \mathbf{\ddot{u}},
\end{equation}
\begin{equation}\label{eq.13}
V^2_S \left( 1 - \ell^2_2 \nabla^2 \right) \nabla^2 \left( \nabla \times \mathbf{u} \right) = \left( 1 - h^2 \nabla^2 \right) \nabla \times \mathbf{\ddot{u}},
\end{equation}
where $V_P = [( \lambda + 2 \mu )/ \rho]^{1/2}$ and $ V_S = \left( \mu/ \rho \right)^{1/2}$ are the propagation velocities of the pressure (P) and shear (S) waves, respectively, in classical elasticity. Moreover, we note that unlike the corresponding case of classical elastodynamics, the partial differential equations \eqref{eq.12} and \eqref{eq.13} are of the fourth order which implies that wave signals emitted from a disturbance point are expected to propagate at different velocities. 

The last statement can be supported by considering a time-harmonic plane wave solution and determining dispersion relations. Specifically, we consider a plane wave solution of Eqs. \eqref{eq.12} and \eqref{eq.13} in the following form
\begin{equation}\label{eq.14}
\mathbf{u} =A \mathbf{d} \exp\Bigl[i \bigl(q \left(\mathbf{n} \cdot \mathbf{x}\right) - \omega t \bigr)\Bigr],
\end{equation}
where $A$ denotes the amplitude, $(\mathbf{d}, \mathbf{n})$ are unit vectors defining the directions of motion and propagation, respectively, $\mathbf{x}$ is the position vector, $q$ is the wavenumber, $\omega$ is the circular frequency of the plane wave, and $i^2=-1$. Then, substituting \eqref{eq.14} into Eqs. \eqref{eq.12} and \eqref{eq.13}, we obtain the following dispersion relations for the pressure and shear waves:
\begin{equation}\label{eq.15}
\omega^2 = V^2_P q^2_P \left( 1 + \ell^2_1 q^2_P \right) \left( 1 + h^2 q^2_P \right)^{-1},
\end{equation}
\begin{equation}\label{eq.16}
\omega^2 = V^2_S q^2_S \left( 1 + \ell^2_2 q^2_S \right) \left( 1 + h^2 q^2_S \right)^{-1},
\end{equation}
where $q_P$ and $q_S$ are the respective wavenumbers. Accordingly, the phase velocities of the pressure $V_P^p$ and shear $V_S^p$ waves in form II of Mindlin's general theory take the following forms:
\begin{equation}\label{eq.17}
V^p_P \equiv \frac{\omega}{q_P} = V_P \left( 1 + \ell^2_1 q^2_P \right)^{1/2} \left( 1 + h^2 q^2_P \right)^{-1/2},
\end{equation}
\begin{equation}\label{eq.18}
V^p_S \equiv \frac{\omega}{q_S} = V_S \left( 1 + \ell^2_2 q^2_S \right)^{1/2} \left( 1 + h^2 q^2_S \right)^{-1/2}.
\end{equation}
Eqs. \eqref{eq.17} and \eqref{eq.18} show that the propagation velocities of these waves depend on the respective wavenumber. Hence, both waves are dispersive in form II of Mindlin's general theory. It is important however to emphasize that the dispersive behavior introduced by the Toupin–Mindlin gradient elasticity is confined to an intermediate range of wavenumbers. Indeed, in the high-wavenumber limit $q \to \infty$, both  velocities in gradient elasticity asymptotically approach a constant value: $V_j (\ell_i/h)$ which, in turn, implies that the medium eventually behaves as effectively non-dispersive at very short wavelengths or high frequencies, but with a renormalized wave speed. This plays a key role in the results presented later where wave trapping and localized resonances are shown to arise in this intermediate dispersive regime where the scale interplay between the wavelength and the microstructural characteristic lengths $\ell$ and $h$ becomes critical. Beyond this range, dispersion effects diminish, and the ability of the structure to sustain confined, resonant modes is substantially reduced. Finally, it is worth noting that in couple stress elasticity only shear waves are dispersive \cite{gourgiotis2016stress}.

To investigate further the nature of the dispersion relations in form II of Mindlin's general theory, we consider the group velocity $V^g = d \omega / d q$. In particular, according to Eqs. \eqref{eq.15}-\eqref{eq.18} we obtain
\begin{equation}\label{eq.19}
V^g_P = V^p_P + \left( \ell^2_1 - h^2 \right) V_P q^2_P \left( 1 + \ell^2_1 q^2_P \right)^{-1/2} \left( 1 + h^2 q^2_P \right)^{-3/2},
\end{equation}
\begin{equation}\label{eq.20}
V^g_S = V^p_S + \left( \ell^2_2 - h^2 \right) V_S q^2_S \left( 1 + \ell^2_2 q^2_S \right)^{-1/2} \left( 1 + h^2 q^2_S \right)^{-3/2}.
\end{equation}
The following three cases are then distinguished: $(i)$ For $\ell^2_i<h^2$, Eqs. \eqref{eq.19} and \eqref{eq.20} imply that $V^g_i<V^p_i$, thus the dispersion is normal. $(ii)$ For $\ell^2_i>h^2$, we have $V^g_i>V^p_i$ indicating that the dispersion is anomalous. $(iii)$ For $\ell_i=h$ or $(\ell_i,h)\rightarrow 0$ (i.e. no material microstructure), the wave velocities degenerate into the non-dispersive velocities of classical elastodynamics \cite{gourgiotis2013reflection}.  

Materials exhibiting normal dispersion typically correspond to classical homogeneous solids without significant microstructural effects such as metals (steel and aluminum), homogeneous glasses, and many polymers at frequencies where their internal structure does not induce complex interactions. On the other hand, anomalous dispersion characterizes materials with pronounced microstructure and internal length scales, such as metamaterials and phononic crystals \cite{srivastava2015elastic}. These engineered materials possess local resonators or periodic heterogeneities that give rise to complex wave phenomena including band gaps, negative refraction, and enhanced localization. Interestingly, it is within this anomalous range that the most intriguing effects emerge in our study including the formation of sharp resonances and strong wave localizations.

We now turn our attention to the energy characteristics of the wave field. The time-average of an energy density field $\mathcal{E}$ over a period, $\mathcal{T} = 2\pi / \omega$, is defined as
\begin{equation}\label{eq:averaged density}
\langle \mathcal{E} \rangle \equiv \frac{1}{\mathcal{T}} \int_t^{t+\mathcal{T}} \mathcal{E} \, dt.
\end{equation}

Using Eqs.~\eqref{eq.1}, \eqref{eq.5}, and \eqref{eq.14}, along with the definitions of $\varepsilon_{ij}$ and $\kappa_{ijk}$, the time-averaged strain and kinetic energy densities over one period $\mathcal{T}$ assume the following form:
\begin{equation}\label{eq:averaged strain energy}
\langle W \rangle= \frac{A^2 q^2}{4} \left\{(\lambda + 2\mu)\beta^2(1+\ell_1^2 q^2)+\mu(1-\beta^2)(1+\ell_2^2 q^2)\right\} \, ,
\end{equation}
\begin{equation}\label{eq:averaged kinetic energy}
\langle T \rangle=\frac{A^2 \rho \omega^2}{4} (1+h^2 q^2) \, ,
\end{equation}
where $\beta = d_i n_i$. Note that in the derivation of Eqs.~\eqref{eq:averaged strain energy} and \eqref{eq:averaged kinetic energy}, the identity 
$\langle \operatorname{Re}(f) \cdot \operatorname{Re}(F) \rangle = \frac{1}{2} \operatorname{Re}(f \cdot \tilde{F})$  has been used, where $f$ and $F$ denote complex time-harmonic fields, and the tilde indicates complex conjugation (see Achenbach~\cite{achenbach2012wave}).

Furthermore, after lengthy but straightforward calculations, the time-averaged power is obtained from Eq.~\eqref{power} as
\begin{equation}
\langle \mathcal{P} \rangle=-\frac{A^2 q \omega}{2} \left\{(\lambda + 2\mu)\beta^2(1+2\ell_1^2 q^2)+\mu(1-\beta^2)(1+2\ell_2^2 q^2)-\rho h^2 \omega^2\right\} \, .
\end{equation}
For (S) waves, the motion is perpendicular to the direction of propagation, implying $\beta = 0$. Applying the dispersion relation \eqref{eq.16}, the time-averaged kinetic and strain energy densities become:
\begin{equation}\label{eq:averaged energy S}
\langle T \rangle = \langle W \rangle = \frac{A^2 q^2 \mu}{4} \left( 1 + \ell^2_2 q^2 \right),
\end{equation}
Moreover, the velocity of the energy transport $c_e$ is given as 
\begin{equation}\label{eq:energyvelS}
c_e=-\frac{\langle \mathcal{P} \rangle}{\langle T \rangle+\langle W \rangle}=-\frac{\langle \mathcal{P} \rangle}{2\langle W \rangle}=\frac{\omega}{q} \left[1+\frac{1}{1+h^2q^2}-\frac{1}{1+\ell_2^2q^2}\right]=V_S^g .
\end{equation}
For (P) waves, the motion and propagation directions are aligned, so that $\beta = 1$. Using the dispersion relation \eqref{eq.15}, we obtain
\begin{equation}\label{eq:averaged energy P}
\langle T \rangle = \langle W \rangle = \frac{A^2 q^2 \left( \lambda + 2 \mu \right)}{4} \left( 1 + \ell^2_1 q^2 \right).
\end{equation}
Consequently, the energy transport velocity for (P) waves becomes
\begin{equation}\label{eq:energyvelP}
c_e=-\frac{\langle \mathcal{P} \rangle}{2\langle W \rangle}=\frac{\omega}{q} \left[1+\frac{1}{1+h^2q^2}-\frac{1}{1+\ell_1^2q^2}\right]=V_P^g\, .
\end{equation}
In both cases of (P) waves and (S) waves the time-averaged kinetic and strain energy densities over a period $\mathcal{T}$ are equal, resulting in a total mechanical energy density of $\langle E \rangle = \langle T \rangle + \langle W \rangle = 2 \langle T \rangle = 2 \langle W \rangle$, as in the case of classical elasticity \cite{rosi2016anisotropic}. Moreover, Eqs \eqref{eq:energyvelS} and \eqref{eq:energyvelP} confirm that, in strain gradient elasticity, the group velocity coincides with the energy transport velocity in a dispersive medium \cite{achenbach2012wave}.

\section{Time-harmonic Green's function in antiplane strain gradient elasticity}
\label{sec3}		
\noindent
For an infinite body occupying the $(x,y)$-plane, with the $z$-axis normal to this plane, and under conditions of antiplane shear, the displacement and body force fields take the following general form:
\begin{equation}\label{eq.21}
u_x \equiv 0, \quad  u_y \equiv 0, \quad u_z \equiv u_z(x,y,t) \neq 0,
\end{equation}
\begin{equation}\label{eq.22}
F_x \equiv 0, \quad  F_y \equiv 0, \quad F_z \equiv F_z(x,y,t) \neq 0.
\end{equation}
All derivatives with respect to $z$ are zero. Under these conditions, the displacement equations of motion of form II \eqref{eq.10} degenerate to a single scalar differential equation \cite{mishuris2012steady}
\begin{equation}\label{PDE}
\mu \left( 1 - \ell^2 \nabla^2 \right) \nabla^2 u_z + \rho h^2 \nabla^2 \ddot{u}_z - \rho \ddot{u}_z + F_z = 0,
\end{equation}
where we set $\ell \equiv \ell_2$, as the microstructural parameter $\ell_1$ does not appear in the displacement equation of motion under antiplane shear conditions. Accordingly, we assume a time harmonic response for the displacement $u_z$ and the body force $F_z$
\begin{equation}\label{eq.24}
u_z(x,y,t) = w(x,y) e^{-i \omega t},
\end{equation}
\begin{equation}\label{eq.25}
F_z(x,y,t) = F_z(x,y) e^{-i \omega t}.
\end{equation}
In view of Eqs. \eqref{eq.24} and \eqref{eq.25}, the equation of motion under time-harmonic conditions takes the form
\begin{equation}\label{eq.26}
\mu \left( 1 - \ell^2 \nabla^2 \right) \nabla^2 w + \rho \omega^2 \left( 1 - h^2 \nabla^2 \right) w + F_z = 0.
\end{equation}
Eq. \eqref{eq.26} suggests that the particle's motion is polarized in the $z$-direction, indicating that the resulting waves are shear-horizontal (SH) waves. Moreover, Eq. \eqref{eq.26} is a non-homogeneous bi-Helmholtz equation, meaning that it can be factorized as the product of a standard Helmholtz operator, associated with propagating wave modes and a modified Helmholtz operator, associated with evanescent modes
\begin{equation}\label{THPDE}
\mu \ell^2 \left( \nabla^2 + q^2_1 \right) \left( \nabla^2 - q^2_2 \right)w = F_z,
\end{equation}
where $(q_1, q_2)$ are real wavenumbers that are related to the solutions of the dispersion equation for the shear waves \eqref{eq.16}:
\begin{equation}\label{eq.28}
q^2_1 = -\frac{1-h^2 k_S^2 }{2\ell^2} +\frac{\sqrt{ 4 \ell^2 k_S^2 + \left( 1-h^2 k_S^2 \right)^2 }}{ 2 \ell^2 },
\end{equation}
\begin{equation}\label{eq.29}
q^2_2 =  \frac{1-h^2 k_S^2 }{2\ell^2} +\frac{\sqrt{ 4 \ell^2 k_S^2 + \left(1- h^2 k_S^2 \right)^2 }}{ 2 \ell^2 }.
\end{equation}
From the above equations, it follows that $\pm q_1$ are the real roots of the dispersion equation \eqref{eq.16}, while $\pm i q_2$ are the imaginary roots, with $k_S = \omega/ V_S$. The signs of $q_1$ and $q_2$ are determined by the sign of $\mathbf{n} \cdot \mathbf{x}$ in Eq. \eqref{eq.14}, which indicates the direction of wave propagation. Specifically, if  $\mathbf{n} \cdot \mathbf{x} > 0$, the wave propagates outwards (in the positive radial direction), meaning it is an outgoing wave, and the evanescent mode decays accordingly. In this case, both $q_1$ and $q_2$ should take positive values. 

Setting $F_z=- \delta(\mathbf{r} - \mathbf{r}')$ in Eq. \eqref{THPDE} where $\delta(\mathbf{r})=\delta(x)\delta(y)$ is the Dirac delta function, we derive the time harmonic isotropic Green's function which gives the out-of-plane displacement at position $\mathbf{r}=(x,y)$, resulting from a concentrated force of unit magnitude acting at a point with position vector $\mathbf{r}' =(x',y')$. In this case, the solution of Eq. \eqref{THPDE} takes the form
\begin{equation}\label{Green}
g(\mathbf{r}; \mathbf{r}') = \frac{1}{4 \mu \pi \ell^2 \left(q^2_1 + q^2_2\right)} \left[ i \pi H^{(1)}_0 (q_1 s) - 2 K_0 (q_2 s) \right],
\end{equation}
where $s = | \mathbf{r} - \mathbf{r}'|$, $H^{(1)}_0$ is the zeroth order Hankel function of the first kind and $K_0$ is the zeroth order modified Bessel function of the second kind. Eq. \eqref{Green} represents the spatial part of the time-harmonic antiplane Green's function. The term $H^{(1)}_0$ combined with the time-harmonic response $e^{- i \omega t}$ corresponds to outgoing waves, whereas the term $K_0$ characterizes exponentially decaying evanescent modes as $s$ increases. An analytical derivation of the Green's function in \eqref{Green} is provided in \hyperref[A]{Appendix~A}.  Note that for an orthotropic medium, the time harmonic antiplane Green's function was previously derived by Gourgiotis and Bigoni \cite{gourgiotis2017dynamics}, but within the framework of couple stress elasticity rather than strain gradient elasticity. Nonetheless, in the case of antiplane deformations, the two theories yield very similar behavior and lead to comparable fundamental solutions.

It is worth noting that unlike the corresponding solution of classical elasticity, which exhibits a logarithmic singularity \cite{graff2012wave}, the Green's function in strain gradient elasticity is bounded in the limit where $ \mathbf{r} \rightarrow \mathbf{r}' $, (i.e., as $s \rightarrow 0$), indicating that the singularity of classical elasticity is now eliminated. Indeed, expanding in series the Green's function as $s \to 0$, we obtain
\begin{equation}\label{eq.32}
g(\mathbf{r}; \mathbf{r}') \sim \frac{1}{\mu \ell^2} \left[ \frac{i \pi + 2 \ln{(q_2/q_1)}}{4 \pi \left(q^2_1 + q^2_2 \right)}\right] + \mathcal{O}(s^2 \ln s).
\end{equation}
In fact, it can be readily shown that the antiplane Green's function is $C^1-$continuous for $s \ge 0$. In particular, we have
\begin{equation}\label{Green first derivative}
\frac{d g}{d s} = \mathcal{O}(s \ln s), \qquad \frac{d^2 g}{d s^2} = \mathcal{O}(\ln s), \qquad \text{as } s \to 0,
\end{equation}
which implies that a time-harmonic concentrated antiplane force induces bounded strains $\varepsilon_{ij}$, but logarithmically unbounded strain gradients $\kappa_{ijk}$ at the origin $s = 0$. This has important implications: the total strain energy (the integral of the strain energy density) remains finite in any region surrounding the concentrated load, illustrating a key regularizing feature of strain gradient theory. This is in marked contrast with classical elasticity, where the strain energy becomes singular and unbounded in the 2D time-harmonic antiplane Kelvin problem.

Finally, it is interesting to note that the displacement equation of motion of antiplane gradient elasticity \eqref{PDE} is identical with the equation that describes the out of plane motion of a prestressed isotropic Kirchhoff plate \cite{reddy2006theory}
\begin{equation}\label{eq.33}
N  \nabla^2 w-D \nabla^4 w + q_z = \rho a \ddot{w}  - \frac{\rho a^3}{12} \nabla^2 \ddot{w}.
\end{equation}
In Eq. \eqref{eq.33} $w \equiv w(x,y)$ now is the deflection, $q_z$ is the distributed load per unit area, $N$ is the isotropic pre-stress in the plate and $D = E a^3/[12 (1 - \nu^2)]$ is the flexural rigidity per unit length, with $E$ being the Young's modulus, $a$ the thickness of the plate and, $\nu$ the Poisson's ratio. This equation was first given by Clebsch (~\cite{clebsch1883theorie}, p.~797, equation (318a)). Comparing  Eqs. \eqref{PDE} and \eqref{eq.33}, it becomes clear that they are essentially the same, differing only in the coefficients of the differential operators. This similarity is useful, as it allows us to interpret the microstructural parameters in terms of more familiar physical quantities. Specifically, we have that $a=\sqrt{12} h$,  $D= \mu \ell^2 a$ and $N=\mu a$ (e.g., \cite{gavardinas2018karman}).

\section{Scattering of SH waves by a cluster of pins}
\label{sec4}		
\begin{figure}[!htb]
\centering
\includegraphics[scale=0.80]{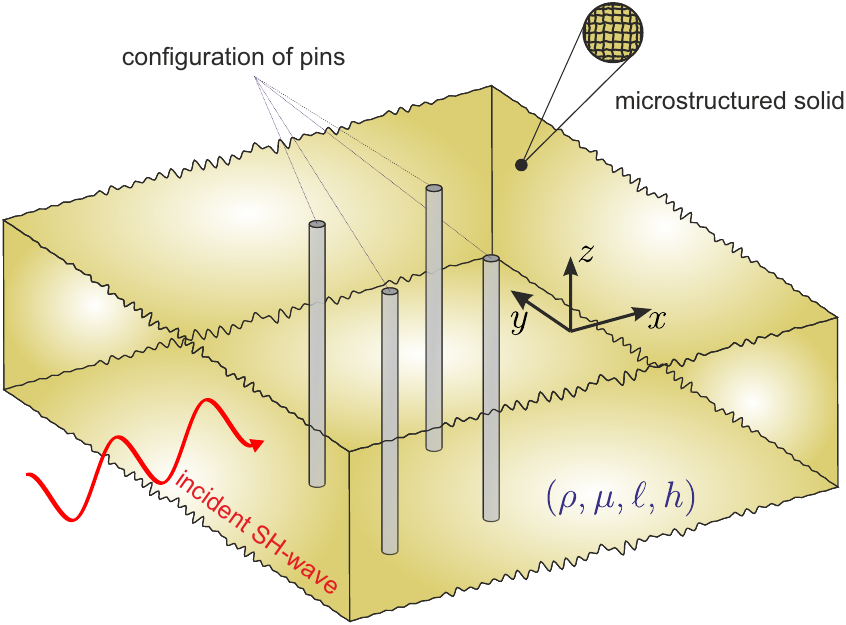}
\caption{Configuration of pins embedded in an infinite elastic microstructured medium subjected to an incident SH harmonic plane wave.}
\label{fig01}
\end{figure}

\noindent
The problem under investigation is illustrated in Fig. \ref{fig01}, which shows antiplane shear (SH) waves propagating through an infinite medium containing a cluster of a finite number of pins. In the antiplane strain setting, the pins are effectively rigid, fibre-like inclusions with zero cross-sectional area along the $z$-axis. An incident SH plane wave of unit amplitude propagates in the $(x,y)$-plane, generating an out-of-plane displacement of the form
\begin{equation}\label{eq.34}
u_z^{\text{(in)}}(x,y,t)=U^\text{(in)}(\mathbf{r}) \cdot e^{- i \omega t},
\end{equation}
with
\begin{equation}\label{eq.35}
U^{\text{(in)}}(\mathbf{r}) = e^{i q_1 \left[x \cos(\psi) + y \sin(\psi) \right]} = e^{i q_1 r \cos(\theta - \psi)},
\end{equation}
where ${\bf r}=(x,y)=(r \cos \theta, r \sin \theta)$,  $q_1 \equiv q_1(\omega)$ is the wavenumber associated with the propagating mode and given as the positive real root of the dispersion equation \eqref{eq.28}, and $\psi$ is the incidence angle between the positive $x$-axis and the direction of wave propagation.

To facilitate the numerical computations, the Green's function \eqref{Green} is expressed using the following dimensionless parameters:
\begin{equation}\label{eq.36}
X = \frac{x}{\ell}, \quad Y = \frac{y}{\ell}, \quad H = \frac{h}{\ell}, \quad K_S = \ell k_S = \frac{\ell \omega}{V_S}.
\end{equation}
This normalisation enables the classification of the dispersion behavior using solely the dimensionless parameter $H$: $(i)$ For $\lvert H \rvert > 1$, Eq. \eqref{eq.20} implies that $V^g_S<V^p_S$, thus the dispersion is normal. $(ii)$ For $\lvert H \rvert < 1$, we have $V^g_S>V^p_S$ indicating that the dispersion is anomalous. $(iii)$ For $\lvert H \rvert = 1$, the wave velocity of the shear waves degenerates into the non-dispersive velocity of classical elastodynamics (see e.g. \cite{gourgiotis2015torsional, zisis2023wave}).

We consider now a cluster of $N_p$ pins at points with position vectors $\mathbf{R}_n'$ = $(X_n',Y_n')$. The total non-dimensional displacement $U(\mathbf{R})$ (i.e. normalized with the unit amplitude of the SH wave) at any point of the infinite domain with position vector $\mathbf{R} = (X,Y)$, can be expressed as the superposition of the displacement generated by the incident wave, given by Eq. \eqref{eq.35}, and the displacement induced by the scattering of the SH waves due to the presence of all the pins. We enforce the null-displacement constraint at the pins by introducing fictitious concentrated body forces at the pin locations. These forces are chosen so that the displacements they generate exactly cancel the displacements produced by the incident SH waves at the pin locations. By the principle of superposition, the total displacement at each pin is therefore zero. The displacements generated by the concentrated body forces are readily determined using the Green's function \eqref{Green}. Thus, the total displacement is given by superposition as

\begin{equation}\label{Disp.Scater}
U(\mathbf{R}) = U^{(\text{in})}(\mathbf{R}) + \sum_{n=1}^{N_p} \left[ A_n \, g(\mathbf{R}; \mathbf{R}'_n) \right],
\end{equation}
where $A_n$ are constants which are obtained by solving the following non-homogeneous system of $N_p$ algebraic equations at the pin locations $\mathbf{R} = \mathbf{R}'_m$ ($m = 1,2,...,N_p$)
\begin{equation}\label{eq.38}
0 = U^{(\text{in})}(\mathbf{R}'_m) + \sum_{n=1}^{N_p} \left[ A_n g(\mathbf{R}'_m; \mathbf{R}'_n) \right] \, , \qquad (m=1,2,...,N_p).
\end{equation}
The solution of the system of equations \eqref{eq.38} can be expressed as the following matrix equation
\begin{equation}\label{eq.39}
\mathbf{A} = - \mathbf{G}^{-1} U^{(\text{in})},
\end{equation}
where $\mathbf{A}$ is an $N_p \times 1$ vector with components $A_m$, $\mathbf{G} \equiv \mathbf{G}(\omega)$ is the $N_p \times N_p$ symmetric Green's matrix with components $G_{mn}=g(\mathbf{R}'_m; \mathbf{R}'_n)$ and $U^{(\text{in})}$ is an $N_p \times 1$ vector with components $U^{(\text{in})}_n$ i.e., the non-dimensional displacement at each pin location, caused by the incident SH wave in an infinite domain with no pins.

Physically, the constants $A_m$ represent the (normalized) reaction forces at the pins and serve to scale the Green's function at each pin location, ensuring that Eq.~\eqref{eq.38} is satisfied. Since the Green's function also captures the influence of each pin across the entire domain, these constants affect the displacement at all other points, as described in Eq.~\eqref{Disp.Scater}.

For a fixed pin configuration and given microstructural ratio $H$, the Green's matrix $\mathbf{G}$ depends solely on the frequency $\omega$ of the incident wave through the dimensionless parameter $K_S$ (defined in Eq.~\eqref{eq.36}). By plotting $\lvert \det(\mathbf{G}) \rvert$ as a function of $K_S$, one can identify local minima corresponding to resonant conditions that is, frequencies that induce large displacement responses in the infinite medium due to scattering. These resonant frequencies occur at points where the first derivative of $\lvert \det(\mathbf{G}) \rvert$ with respect to $K_S$ vanishes and the second derivative is positive. However, true resonances correspond to \emph{higher-order} local minima where, in addition to the first and second derivatives being, respectively, zero and positive, the third derivative also vanishes and the fourth derivative is negative. This higher-order flatness indicates a sharp, symmetric minimum, signalling near-singularity of $\mathbf{G}$ and resulting in pronounced amplification of the solution amplitudes $A_m$. Accurate detection of these higher-order minima is essential for correctly identifying resonant frequencies. To this end, the Jacobi formula is employed to compute the derivatives of $\det(\mathbf{G})$ precisely.

Using the logarithmic measure
\begin{equation}\label{eq.Gamma}
\mathbb{\Gamma} = \log\bigl(\lvert \det(\mathbf{G}) \rvert\bigr),
\end{equation}
which enhances the contrast between sharp and shallow minima for more reliable detection, these resonance conditions read
\begin{equation}\label{eq.40}
\frac{d \Gamma}{dK_{S}} = 0, \quad 
\frac{d^{2} \Gamma}{dK_{S}^{2}} > 0, \quad 
\frac{d^{3} \Gamma}{dK_{S}^{3}} = 0, \quad 
\frac{d^{4} \Gamma}{dK_{S}^{4}} < 0.
\end{equation}

It is important to emphasize that the term ``resonance'' is used here in a generalized sense. Unlike classical eigenvalue problems, where resonance is associated with non-trivial solutions of a homogeneous system, the present formulation (Eq.~\eqref{eq.38}) deals with a non-homogeneous system. Here, resonance refers to frequencies and wave numbers that maximize the displacement response in the medium due to scattering effects.

\begin{figure}[!htb]
\centering
\includegraphics[scale=0.55]{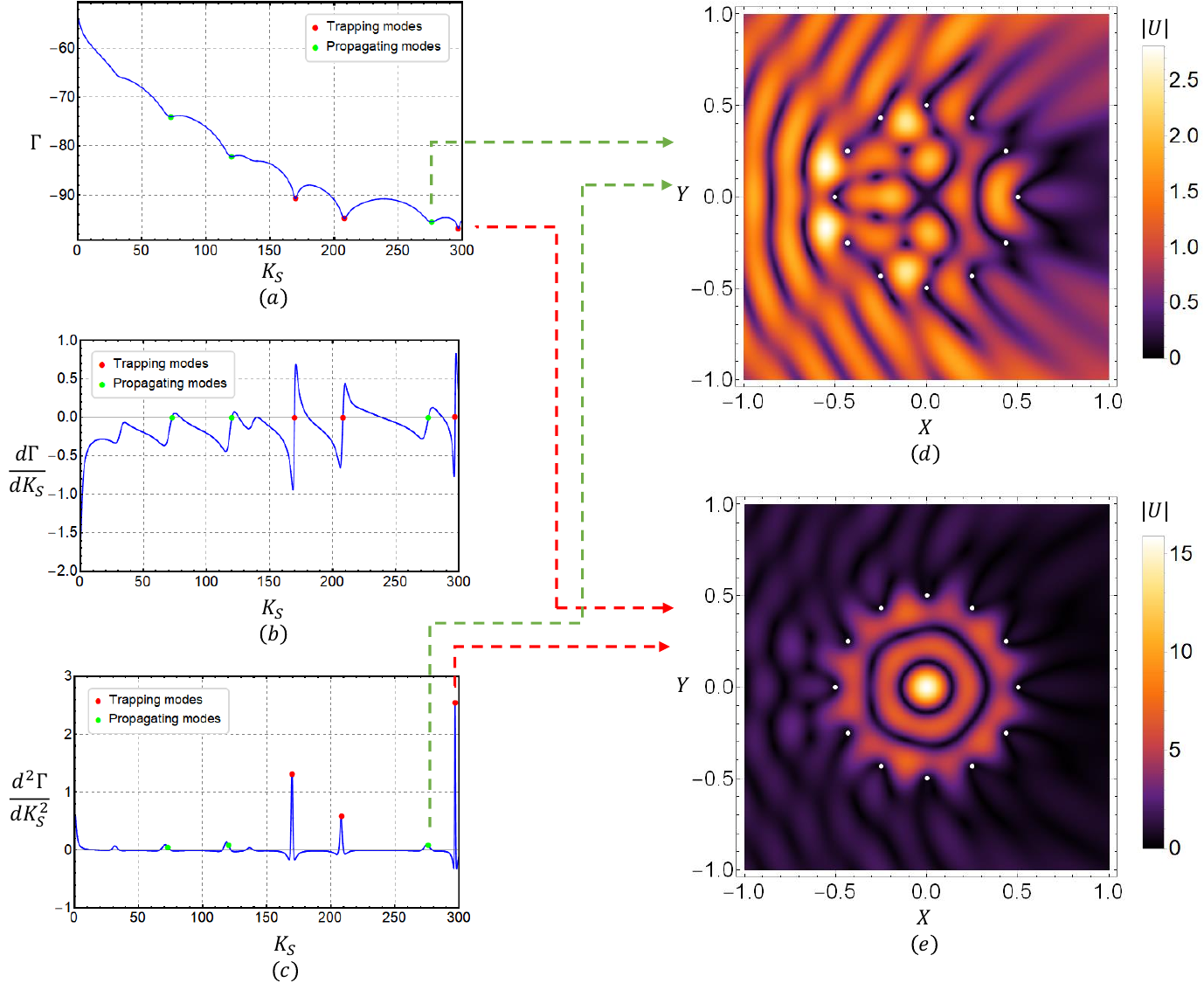}
\caption{$(a)-(c)$ detected resonant frequencies, $(d)$ propagating (non-localized) mode $K_S = 275.851$, and $(e)$ trapping (localized) mode $K_S = 296.959$, for a circular configuration of unit diameter consisting of $12$ pins, $H =0$ (no microinertia).}
\label{fig02}
\end{figure}

Figure~\ref{fig02}a illustrates the detected local minima for a circular arrangement of $12$ pins subjected to unit amplitude SH waves for values of $K_S < 100$. In general, sharper local minima of $\lvert \det(\mathbf{G}) \rvert$ (or equivalently of $\Gamma$) correspond to resonant modes that produce greater displacements in the body. These sharp minima (marked in red) are characterized by a peak in the second derivative of $\Gamma$ with respect to $K_S$ and satisfy the higher-order conditions listed in Eq.~\eqref{eq.40}. In such cases, the wave disturbance remains predominantly confined within the pin cluster. This behavior defines the trapping (localized) modes. In contrast, for other local minima of $\lvert \det(\mathbf{G}) \rvert$ (marked in green), the disturbance does not localize; instead, SH waves propagate through the pin array, producing larger displacements outside the cluster. The same behavior is observed, of course, for random frequencies. This correlation between the sharpness of local minima and the nature of resonant modes holds consistently across all pin configurations considered in this study.

In Figures (\ref{fig02}d) and (\ref{fig02}e) the distribution of the absolute value of the non-dimensional displacement $\lvert U \rvert$ for a circular configuration of unit diameter consisting of $12$ pins under different types of resonances is presented. Fig. (\ref{fig02}d) displays the formation of distinct propagation paths extending beyond the pin array. This corresponds to the $6^{th}$ local minimum point in Fig. (\ref{fig02}a), which does not correspond to a local maximum in $d^2 \Gamma / dK_S^2$, suggesting a non-localized propagative behavior rather than a true resonance. In contrast, in Fig. (\ref{fig02}e) the motion is mainly confined within the configuration of pins. This corresponds to the $7^{th}$ local minimum point in Fig. (\ref{fig02}a), which aligns with a local maximum point of $d^2 \Gamma /dK_S^2$ as shown in Fig. (\ref{fig02}c), indicating a trapping mode. The analysis further indicates that maximum absolute displacements are consistently higher for trapping modes compared to these propagating responses, underscoring the greater ability of the former to localize energy within the pin configuration.

\section{Scattering by Koch's snowflake}
\label{sec5}		
\noindent
Based on the preceding analysis, it is possible to model the scattering of SH waves at any frequency in an infinite domain with a finite number of rigid pins. The configuration of choice in the present work is the Koch snowflake, a fractal curve generated through an iterative process starting from an equilateral triangle \cite{koch1904courbe}. For comparison, we also consider an alternative simple configuration in which the pins are arranged uniformly along the perimeter of a circle of unit diameter as was examined in \hyperref[sec4]{Section~4}. The normalized diameter of the circle is set equal to the side length of the initial triangle used to generate the Koch snowflake, aiming that both configurations share a comparable characteristic length scale. To ensure comparability between the two configurations, the number of pins in each circular arrangement was chosen to match the number of corners of the Koch snowflake in each iteration.

\begin{figure}[!htb]
\centering
\includegraphics[scale=0.40]{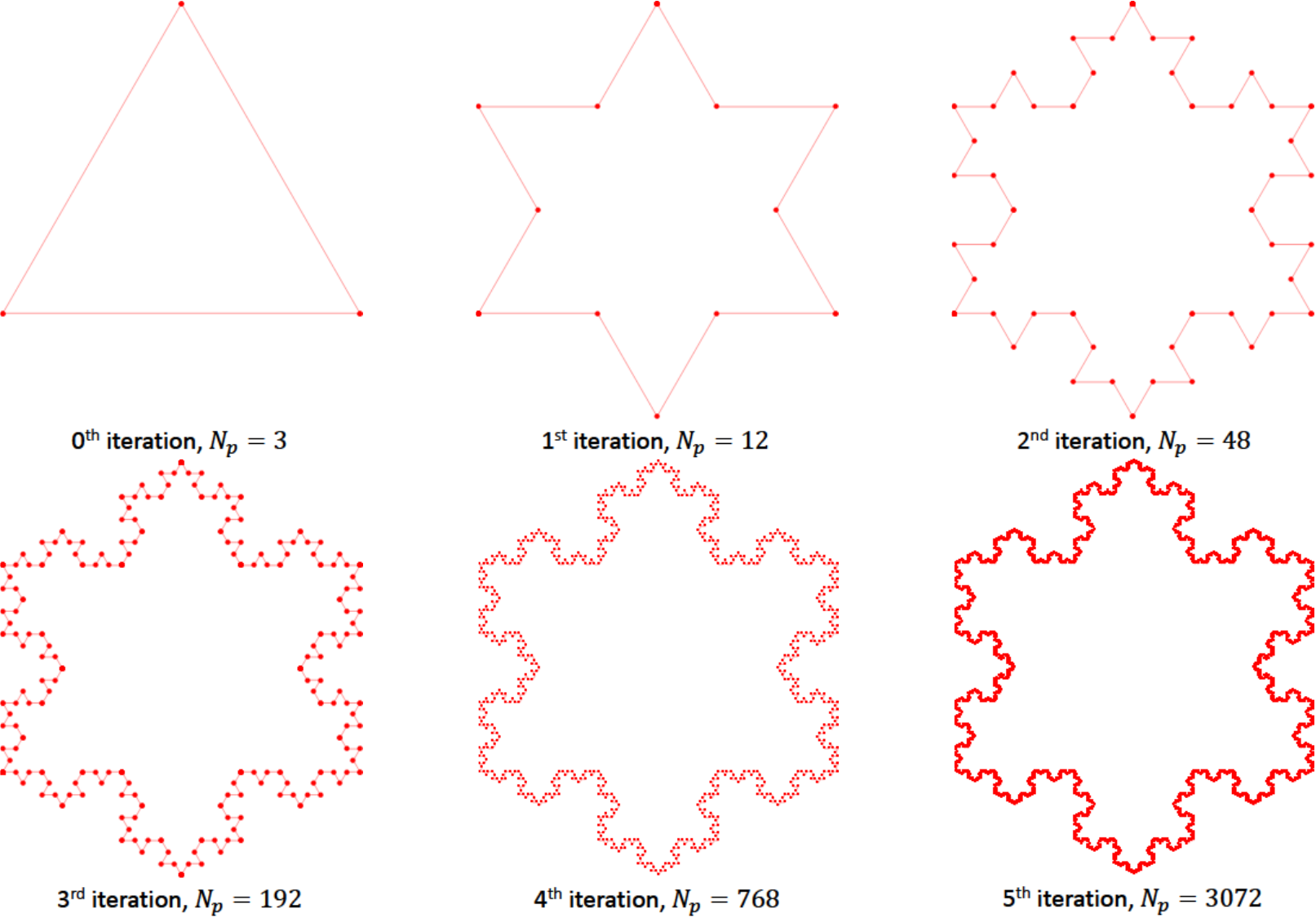}
\caption{First five iterations of the Koch snowflake, $N_{p}$ denotes the number of corners where the pins are located.}
\label{fig03}
\end{figure}

The Koch snowflake is constructed in a sequence of stages starting with an equilateral triangle, followed by a recursive alteration of each line segment as follows: Initially the straight segments are divided into three segments of equal length. Then, an outward pointing equilateral triangle that has the middle segment of the previous iteration as its base is drawn and finally the line segment that coincides with the base of the triangle is removed. The first five iterations of the Koch snowflake are illustrated in Fig. \ref{fig03}, while its geometric properties are presented in \hyperref[B]{Appendix~B}.

\subsection{The effect of dispersion type}
\noindent The analysis shows that the type of dispersion, characterized by the non-dimensional parameter $H$, significantly influences the behavior of $|\det(\mathbf{G})|$ and, consequently, the system's response. The first key observation is that distinct, sharp local minima satisfying the relevant resonance conditions \eqref{eq.40} — and leading to trapping modes — are consistently observed when the dispersion is anomalous (i.e., $H < 1$). This is a geometry-independent feature of the system and hence does not depend on the specific arrangement of the pins. 

\begin{figure}[H]
\centering
\includegraphics[scale=0.80]{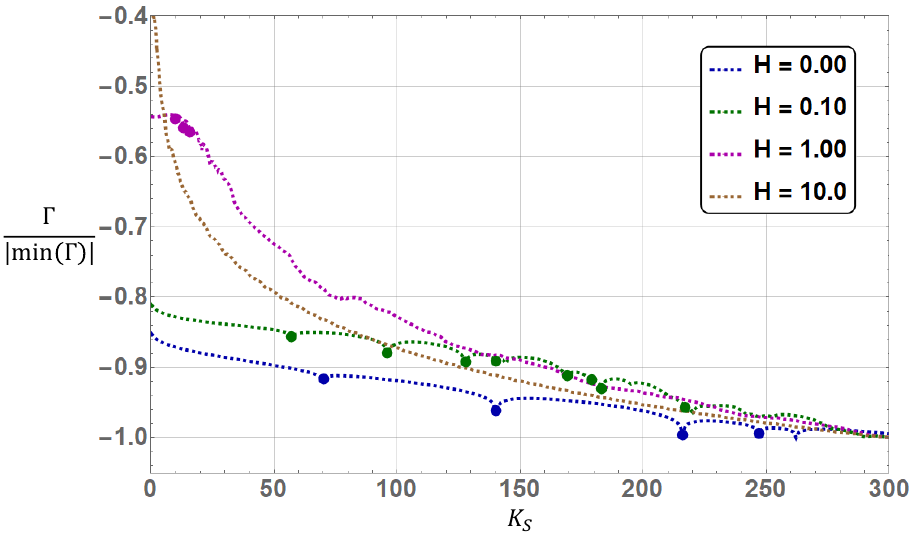}
\caption{Variation of the normalized determinant of the Green's matrix for a pin configuration corresponding to the $2^{nd}$ iteration of the Koch snowflake, for different values of $H$.}.
\label{fig04}
\end{figure}

As shown in Fig.~\ref{fig04}, in the Koch snowflake configuration (2nd iteration - 48 pins), distinct sharp local minima become especially pronounced for $H < 1$ (anomalous dispersion), and in particular for $H = 0$ (zero micro-inertia). These minima satisfy the resonance conditions and are associated with strongly localized trapping modes. As $H$ increases, the number of local minima in $|\det(\mathbf{G})|$ also increases. However, these minima lose their sharpness and no longer correspond to strong resonances. Beyond a certain threshold, $|\det(\mathbf{G})|$ becomes nearly monotonic, and the system response transitions to a delocalized regime. In the case of normal dispersion ($H > 1$), a closer look at Fig.~\ref{fig04} reveals an oscillatory behavior in $|\det(\mathbf{G})|$ characterized by a series of closely spaced local minima with mollified peaks. These features indicate a weakening of resonance effects and a shift toward smoother, less localized dynamics. Similar observations are made when the pins are arranged along the perimeter of a circle, suggesting that the qualitative influence of the dispersion parameter $H$ on the structure of $|\det(\mathbf{G})|$ is largely independent of the pin configuration.

\begin{figure}[!htb]
\centering
\includegraphics[scale=0.38]{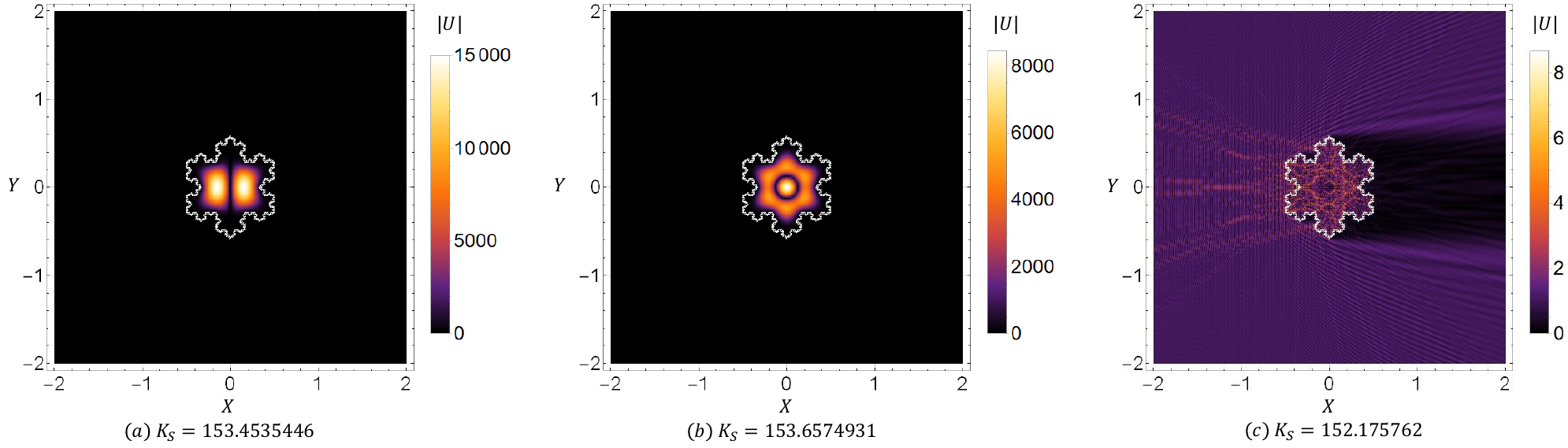}
\caption{Distribution of the non-dimensional displacement for a pined configuration corresponding to the $4^{th}$ iteration of the Koch snowflake: $(a)$ $H = 0$, $(b)$ $H = 0.1$, $(c)$ $H = 1$.}
\label{fig05}
\end{figure}

Fig.~\ref{fig05} presents the distribution of the non-dimensional modulus of displacement for the fourth iteration of the Koch snowflake (768 pins), evaluated at three characteristic values of $H$, each corresponding to a resonance occurring at very similar wavenumbers $K_S$. The results clearly show that the largest displacements consistently occur when $H = 0$, corresponding to the case of zero micro-inertia. This behavior is universal and independent of the pin configuration, as the same trend is also observed for circular arrangements. The implication is that micro-inertia acts as a shielding mechanism, effectively increasing the apparent stiffness of the medium and thereby suppressing the amplitude of resonant motion.

\noindent Moreover, Fig.~\ref{fig05} illustrates how drastically the localization pattern changes with only slight perturbations in $H$, even when the resonant wavenumber remains nearly unchanged. In the non-dispersive case ($H = 1$), the maximum displacement amplitude is reduced by approximately three orders of magnitude compared to the $H = 0$ case and the displacement field is effectively suppressed inside the configuration. This highlights the sensitivity of the resonance and localization mechanisms to the dispersive character of the medium.

\subsection{The effect of pin configuration dimension}
\noindent For a given pin configuration, whether circular or based on the Koch snowflake, the geometric boundary dimension plays a critical role in shaping the wave behavior. In the Koch snowflake case, this dimension evolves with each iteration and can be quantified using the Minkowski-Bouligand fractal (box-counting) dimension (see \hyperref[B]{Appendix~B}). In contrast, for circular configurations, the boundary dimension remains fixed at one regardless of the number of pins.

Table \ref{tab:dimensions} provides the boundary dimensions for the first four iterations of the Koch snowflake, alongside those of circular configurations consisting of the same number of pins.

\begin{table}[h]
    \centering
    \caption{Values of the boundary dimension for the first four iterations of the Koch snowflake and for circular configurations with the same number of pins.}
    \label{tab:dimensions}
    \begin{tabular}{|c|c|c|}
        \hline
        \multirow{2}{*}{\parbox[c]{3cm}{\centering Number of pins \\ $N_p$}} 
        & $\dim$ & $\dim$ \\
        \cline{2-3}
        & Koch snowflake & Circular configurations \\
        \hline
        $12$  & $2.26$ & $1.00$ \\
        $48$  & $1.76$ & $1.00$ \\
        $192$ & $1.60$ & $1.00$ \\
        $768$ & $1.51$ & $1.00$ \\
        \hline
    \end{tabular}
\end{table}
To investigate the impact of the dimension of the configuration of pins, $\Gamma$ is plotted as a function of the non-dimensional wavenumber $K_S$ for the first four iterations of the Koch snowflake as well as for circular configurations with the same number of pins. It is noted that no local minima appear in the zeroth iteration and hence no results are shown for this case.
\begin{figure}[!htb]
\centering
\includegraphics[scale=0.38]{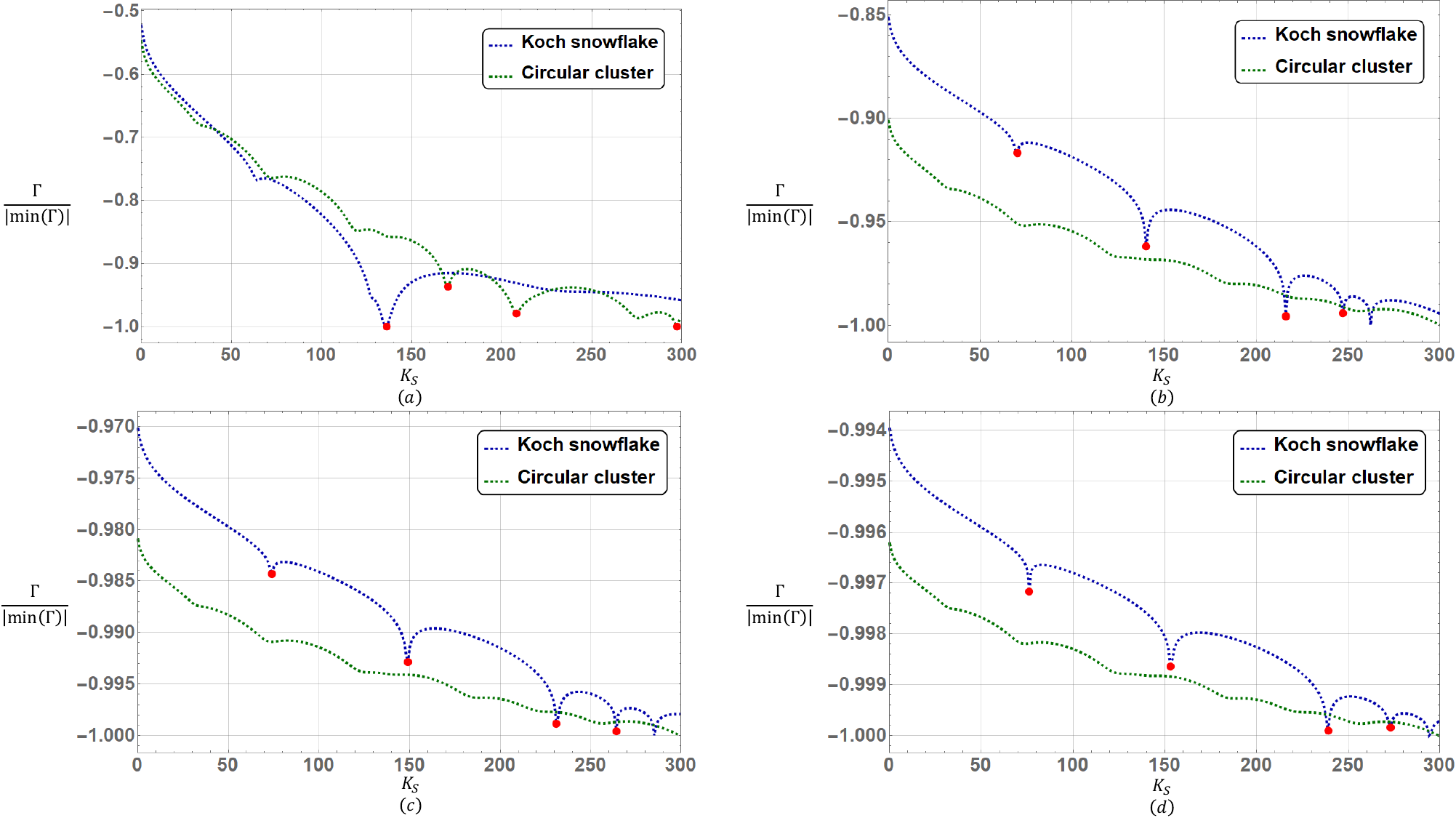}
\caption{Variation of the normalized determinant of the Green's matrix and detected trapping (localized) modes for the Koch snowflake and for circular configurations with: $(a)$ $12$, $(b)$ $48$, $(c)$ $192$, $(d)$ $768$ pins, $H = 0$.}
\label{fig06}
\end{figure}
As shown in Fig. \ref{fig06}, with each additional iteration of the Koch snowflake, the positions of the local minima in $|\det(\mathbf{G})|$ tend to stabilize converging to specific values. Beyond a certain iteration, the number of distinct local minima remains essentially constant regardless of how many pins are inserted. This behavior suggests that it is the geometry and dimension of the configuration and not simply the number of pins that governs the resonant characteristics of the system.

A key difference between the circular and fractal configurations lies in the way that the number and sharpness of resonances evolve as the number of pins increases. For the Koch snowflake, transitioning from the first to the second iteration (i.e., increasing from 12 to 48 pins) leads to an increase of the number of resonances with sharper local minima, as shown in Fig. \ref{fig06}. Beyond a certain iteration, however, the response converges: both the number and the positions of the local minima stabilize, and the graph of $|\det(\mathbf{G})|$ becomes effectively invariant to further increases in pin count (see Fig. \ref{fig06}). In contrast, for circular configurations, with the same number of pins as the corresponding Koch iterations, increasing the pin count from 12 to 48 actually reduces the number of resonances, and the local minima become progressively smoother. Notably, for the circular cases with 48, 192, and 768 pins, no sharp local minima (indicating resonant modes) are observed in the range $0 < K_S < 300$. Only for higher wavenumbers ($K_S > 300$) do resonant features re-emerge, confirming the diminished trapping capability of circular configurations at lower spectral ranges.

\begin{figure}[!htb]
\centering
\includegraphics[scale=0.55]{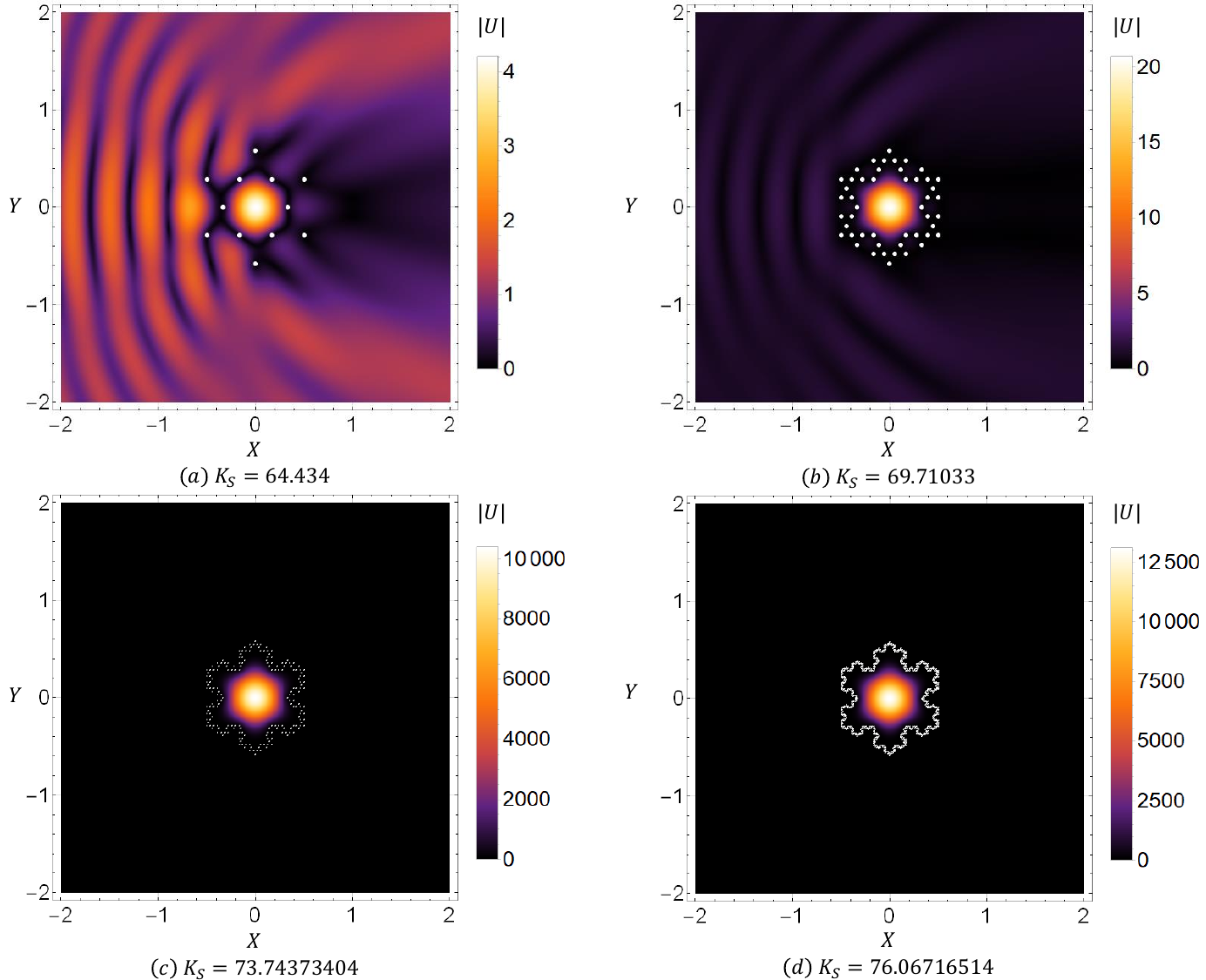}
\caption{Distribution of the non-dimensional displacement for the first four iterations of the Koch snowflake at the first resonant mode.}
\label{fig07}
\end{figure}

\begin{figure}[!htb]
\centering
\includegraphics[scale=0.38]{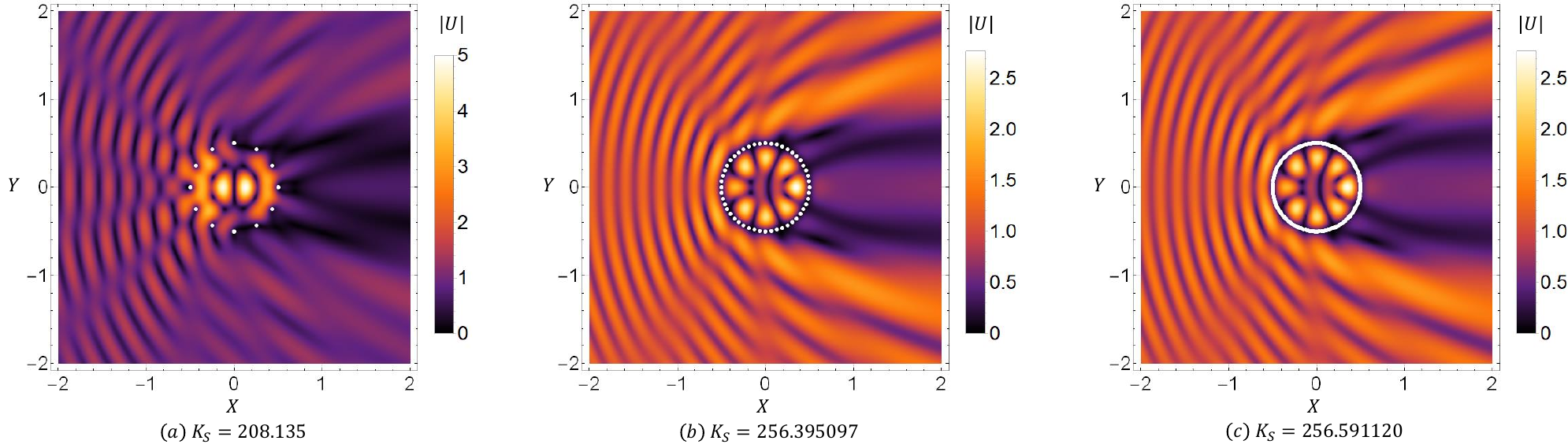}
\caption{Distribution of the non-dimensional displacement for circular configurations with: $(a)$ $12$, $(b)$ $48$, $(c)$ $192$ pins at the first resonant mode, $H = 0$.}
\label{fig08}
\end{figure}

The disparity in the sharpness of the local minima between the two configurations is also reflected in the values of the non-dimensional displacements. For the Koch snowflake, each new iteration results in a noticeable increase in the amplitude of the non-dimensional displacement, as shown in Fig. \ref{fig07}. This reflects the growing ability of the increasingly intricate boundary to trap and localize wave energy. However, this effect saturates beyond the third iteration. By contrast, circular configurations display little change in displacement distribution as the number of pins increases (see Fig.~\ref{fig08}). Since their boundary dimension remains constant and equal to one, the system's behavior is essentially unaffected by the increase in geometric detail. In fact, a slight decrease in displacement amplitude is observed with more pins, reinforcing the conclusion that it is the geometry's dimension, rather than the pin count, that governs wave localization.

\begin{figure}[!htb]
\centering
\includegraphics[scale=0.50]{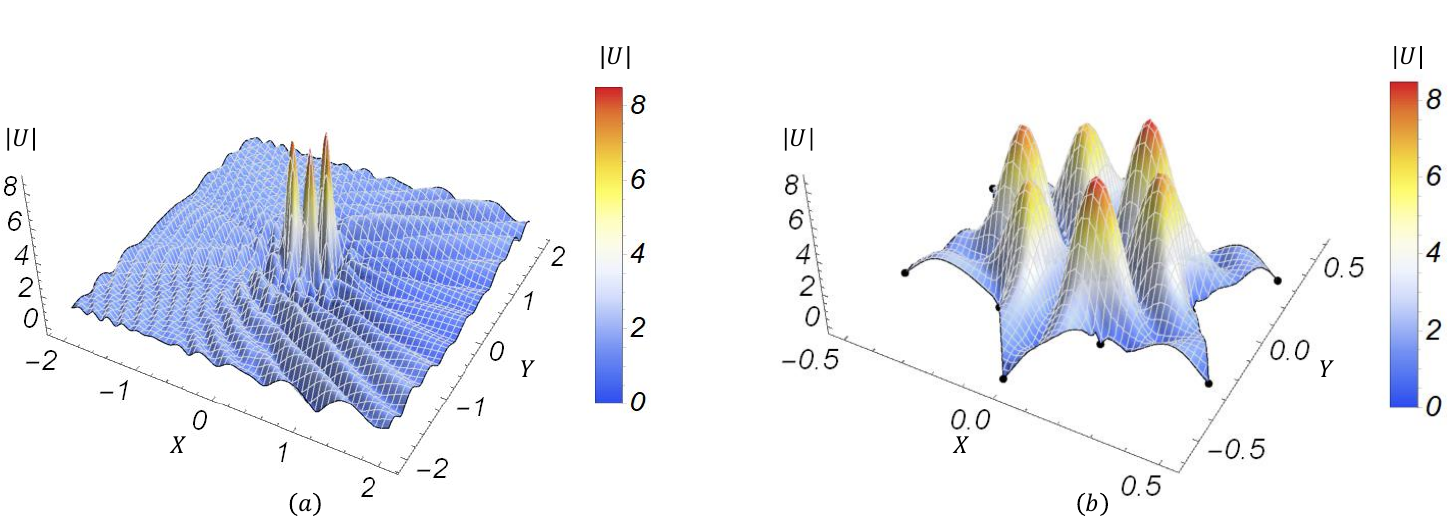}
\caption{Deformed state of the: $(a)$ surrounding area and $(b)$ Koch snowflake, at a trapping (localized) mode, $K_S = 381.592$, $H = 0$, $N_P =12$.}
\label{fig9}
\end{figure}

\begin{figure}[!htb]
\centering
\includegraphics[scale=0.50]{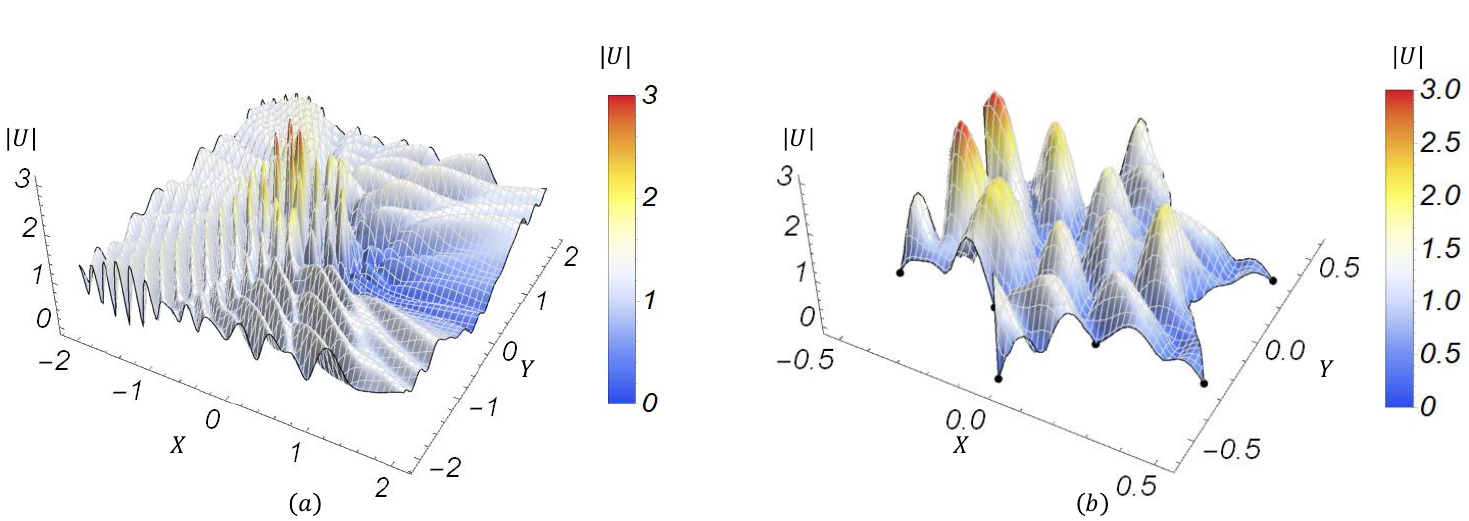}
\caption{Deformed state of the: $(a)$ surrounding area and $(b)$ Koch snowflake, at a propagating (non-localized) response, $K_S = 524.125$, $H = 0$, $N_P = 12$.}
\label{fig10}
\end{figure}

Figs. \ref{fig9} and \ref{fig10} provide a three-dimensional visualization of the deformed state of the snowflake and its surrounding region, for resonant modes corresponding to either wave trapping or propagation. These figures reveal that the wave amplitude is attenuated and the wavelength increases behind the pinned configuration. A similar response, characterized by wave attenuation and wavelength stretching, was observed across all other resonant modes identified, including those involving circular pin arrangements.

\subsection{The effect of the wave incident angle}

\begin{figure}[!htb]
\centering
\includegraphics[scale=0.38]{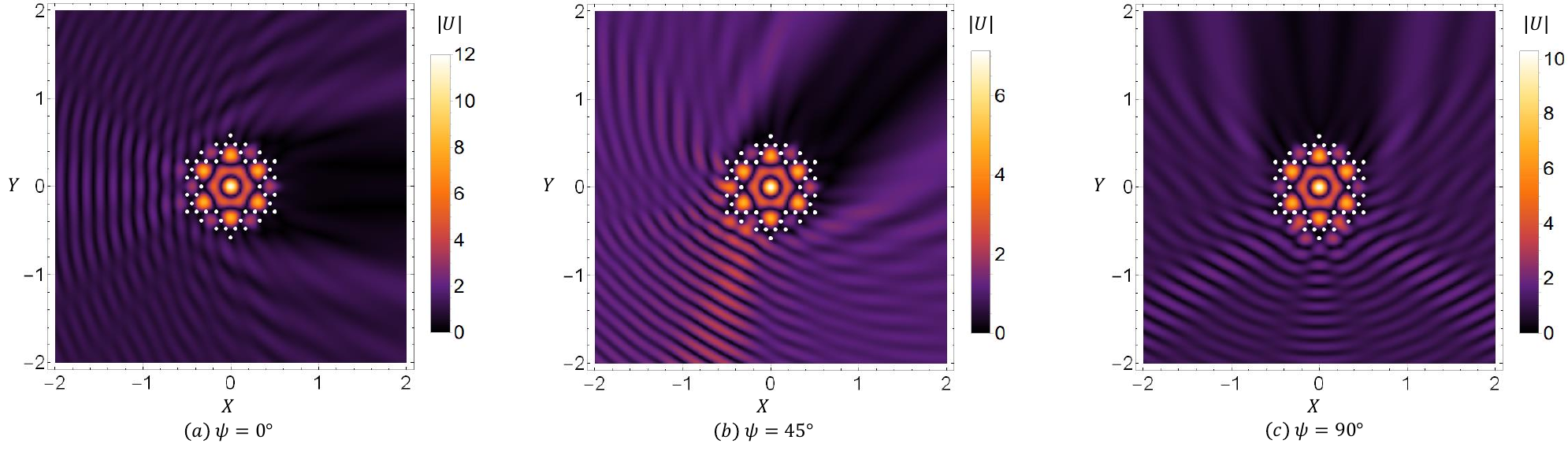}
\caption{Distribution of the non-dimensional displacement for a pined configuration corresponding to the $2^{nd}$ iteration of the Koch snowflake at different incident wave angles when $K_S = 436.43373$, $H = 0$.}
\label{fig11}
\end{figure}

\noindent
We now examine the influence of the angle of incidence $\psi$ of the incoming SH plane waves. The results reveal that $\psi$ plays a crucial role in determining both the scattering patterns and the degree of wave localization. However, the quantity $\lvert \det(\mathbf{G}) \rvert$ remains invariant with respect to $\psi$, as the Green's matrix $\mathbf{G}$ depends solely on the frequency and the spatial arrangement of the pins. Indeed, the incident angle affects only the incoming wave (i.e. the force vector) and not the system matrix itself. As a result, while the resonant frequencies remain fixed, the response amplitude quantified by $\lvert U_{\max} \rvert$ varies with $\psi$. For each resonance, there exists an optimal angle of incidence that maximizes the amplitude. In fact, maximum displacements are observed when the direction of wave propagation aligns with an axis of symmetry in the pin configuration. Among these symmetry directions, the largest responses occur when the incident SH wave impinges directly on a pin located at a non-convex corner relative to its neighboring pins, as shown in Figs.~\ref{fig11}a and \ref{fig11}c. Furthermore, the displacement field exhibits invariance under incident angles $\psi = 0^\circ$, $120^\circ$, $240^\circ$, and $360^\circ$, consistent with the rotational symmetry of the Koch snowflake. Since the geometry possesses threefold rotational symmetry, the physical response naturally repeats every $120^\circ$ rotation~\cite{conway2016symmetries}.

In the case of circular pin arrangements, the angle of incidence, as expected, has minimal effect upon the system's response, particularly when a large number of pins are present. This is of course attributed to the radial symmetry of the circular configuration, which renders the system almost invariant to the direction of wave propagation.   

\subsection{The effect of multiple snowflakes}
\begin{figure}[!htb]
\centering
\includegraphics[scale=0.60]{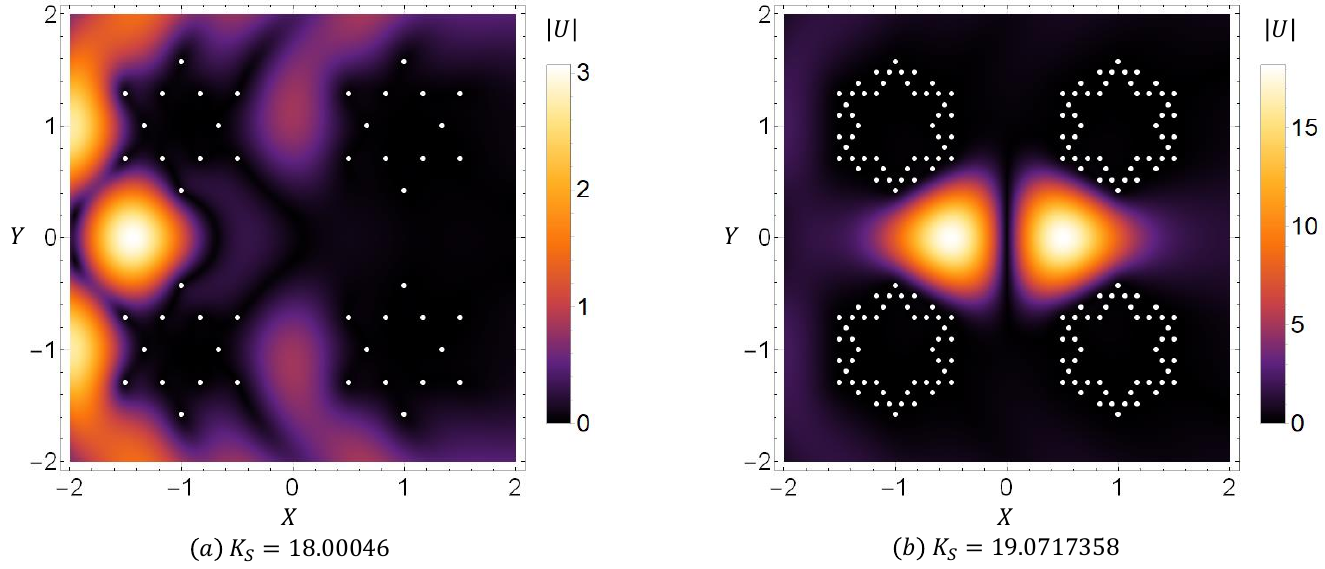}
\caption{Distribution of the non-dimensional displacement for multiple pin configurations corresponding to the: $(a)$ $1^{st}$, $(b)$ $2^{nd}$ iterations of the Koch snowflake at the first resonant mode.}
\label{fig12}
\end{figure}

\noindent Furthermore, we examine the system's response when multiple Koch snowflake configurations are arranged in close proximity to form a compound structure, as shown in Fig. \ref{fig12}. Specifically, we consider four identical snowflakes placed symmetrically to investigate whether the resonant trapping seen in a single snowflake persists within each snowflake or shifts to the spaces between them. Interestingly, in some resonant modes, the central region enclosed by the four snowflakes shows strong wave trapping, with amplified particle motion mostly occurring outside the single pin configurations but inside this enclosed area. This behavior contrasts sharply with the single snowflake or circular configurations, where wave localization is confined within the pin layout.

\begin{figure}[!htb]
\centering
\includegraphics[scale=0.60]{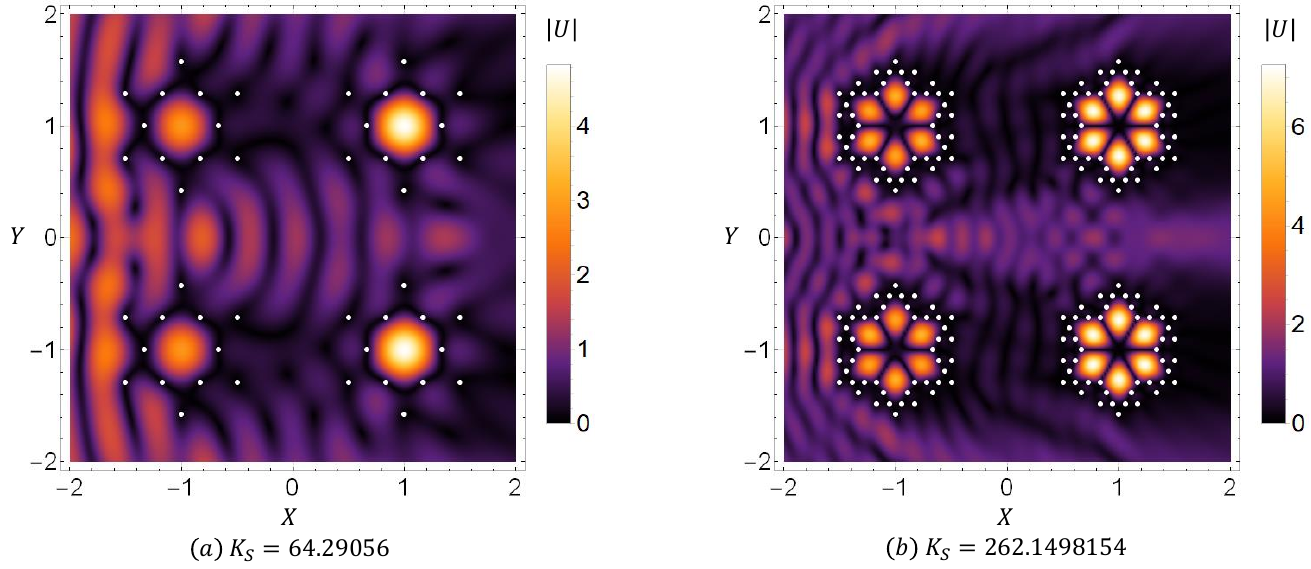}
\caption{Distribution of the non-dimensional displacement for multiple pin configurations corresponding to the: $(a)$ $1^{st}$, $(b)$ $2^{nd}$ iteration of the Koch snowflake, shown at different resonant modes in which waves localise within each pin arrangement.}
\label{fig13}
\end{figure}

This response is attained for relatively small resonant wavenumbers. At these wavenumbers, the corresponding wavelength is large compared to the distance between successive pins, which promotes constructive interference outside the snowflake configurations while suppressing wave penetration into the enclosed area. As the resonant wavenumbers increase, the associated wavelength decreases. Beyond a certain point, the wavelength becomes comparable to or smaller than the spacing between successive pins, allowing waves to propagate into and through the pin configurations. This transition, where the trapping within the enclosed area weakens and resonant modes begin to localize within the individual pin clusters themselves, is illustrated in Fig. \ref{fig13}.

The observed phenomena result from the interplay of scattering, diffraction, and interference. At small resonant wavenumbers, the pins act as strong scatterers, forming partial barriers: destructive interference inside the pin clusters suppresses wave penetration, while constructive interference between clusters reinforces wave confinement in the enclosed region. As frequency rises, the pins become more transparent, allowing waves to partially transmit and weakening this confinement. This transition underscores how geometry and frequency govern resonant wave behavior in complex pin arrangements.   

\subsection{Trapping of mechanical energy}
\noindent
Finally, we present the spatial distributions of time-averaged kinetic, strain, and total mechanical energy densities which are crucial for understanding the dynamic behavior of wave propagation in media with localized constraints, such as pins. Such distributions reveal how the pin arrangements influence the trapping and transfer of wave energy throughout the system.

Considering Eqs. \eqref{eq.21} and the definitions of $\varepsilon_{ij}$ and $\kappa_{ijk}$, along with Eqs. \eqref{eq.11}, the expressions for the kinetic and strain energy densities, originally given in Eqs. \eqref{eq.1} and \eqref{eq.5} can be reformulated in the antiplane strain case as follows
\begin{equation}\label{eq.41}
T = \frac{1}{2} \rho \dot{u}_z^2
+ \frac{1}{2} \rho h^2 \left( \dot{u}_{z,x}^2 + \dot{u}_{z,y}^2 \right),
\end{equation}
\begin{equation}\label{eq.42}
W = \frac{1}{2} \mu \left( u_{z,x}^2 + u_{z,y}^2 \right)
+ \frac{1}{2} \mu \ell^2 \left( u_{z,xx}^2 + 2 u_{z,xy}^2 + u_{z,yy}^2 \right).
\end{equation}
It is important to note that the derivation of Eq.~\eqref{eq.42} relies on the assumption that $\hat{a}_3 = 0$, which is the case of the simplified theory of antiplane strain gradient elasticity. Given the structural analogy between plate bending and antiplane gradient elasticity, this assumption is formally tantamount to setting the Poisson's ratio equal to one ($\nu = 1$) in the thin plate model, i.e. the contribution of the Gaussian curvature to the bending energy vanishes. In this sense, neglecting the $\hat{a}_3$ term corresponds to ignoring Gaussian curvature effects in the energy, retaining only the mean curvature (Laplacian-type) contribution.

The kinetic and strain energy densities, denoted by $\langle T \rangle$ and $\langle W \rangle$ respectively, have their time-averaged values given by the following expressions:
\begin{equation}\label{eq.44}
\langle T \rangle = \frac{1}{4} \mu k_S^2 \lvert w \rvert^2
+ \frac{1}{4} \mu h^2 k_S^2 \left( \lvert w_{,x} \rvert^2 + \lvert w_{,y} \rvert^2 \right),
\end{equation}
\begin{equation}\label{eq.45}
\langle W \rangle = \frac{1}{4} \mu \left( \lvert w_{,x} \rvert^2 + \lvert w_{,y} \rvert^2 \right)
+ \frac{1}{4} \mu \ell^2 \left( \lvert w_{,xx} \rvert^2 + 2 \lvert w_{,xy} \rvert^2 + \lvert w_{,yy} \rvert^2 \right).
\end{equation}
In the derivation of Eqs. \eqref{eq.44} and \eqref{eq.45}, Eqs. \eqref{eq:averaged density} and \eqref{eq.24} have been taken into account together with the properties of the time average of products (see Achenbach \cite{achenbach2012wave}). Since the displacement field $u_z$ is complex-valued in the context of time-harmonic solutions, the notation $\vert \cdot \rvert^2$ refers to the squared modulus, i.e., the product with its complex conjugate. Although for a plane wave $\langle W \rangle=\langle T \rangle$, as shown in \hyperref[sec2]{Section~2}, in the scattering problem under investigation, this equality no longer holds due to energy localization and interference effects induced by the pin configuration.

\begin{figure}[!htb]
\centering
\includegraphics[scale=0.40]{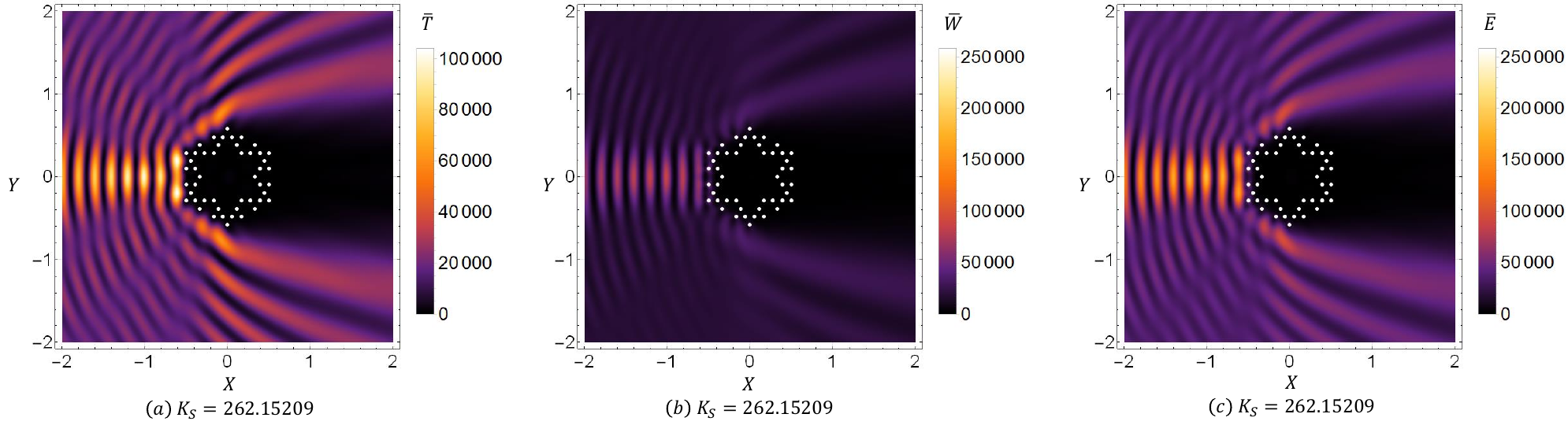}
\caption{Distribution of the non-dimensional time-averaged energy densities: (a) kinetic $\bar{T}$, (b) strain $\bar{W}$, and (c) total mechanical $\bar{E}$ energy for the $2^{nd}$ iteration of the Koch snowflake, for a propagating (non-localized) response: $H = 0$.}
\label{fig14}
\end{figure}

\begin{figure}[!htb]
\centering
\includegraphics[scale=0.40]{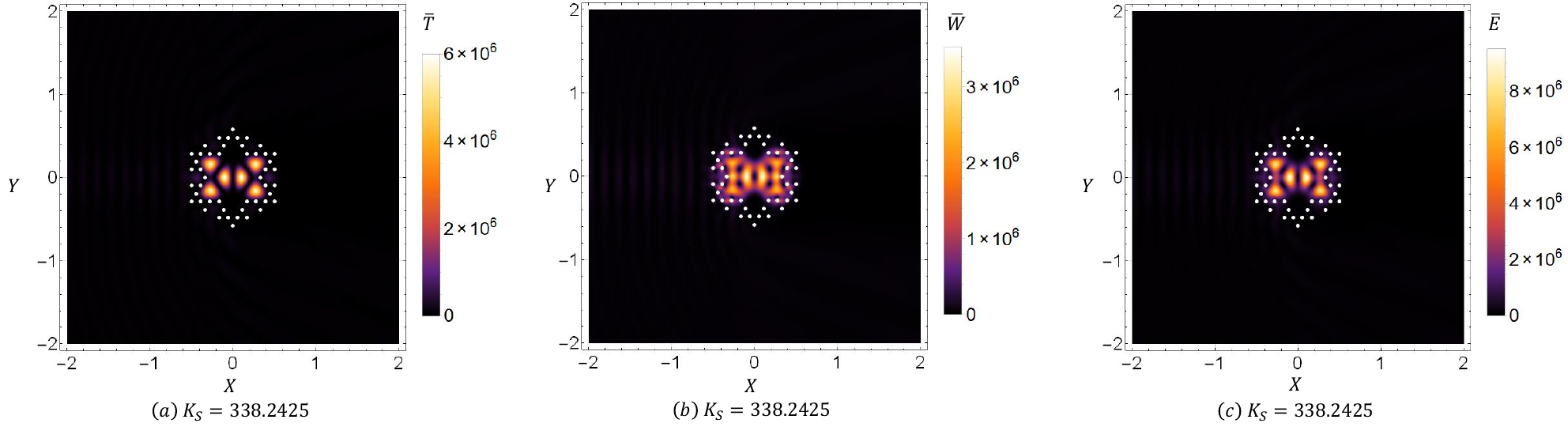}
\caption{Distribution of the non-dimensional time-averaged energy densities: (a) kinetic $\bar{T}$, (b) strain $\bar{W}$, and (c) total mechanical $\bar{E}$ energy for the $2^{nd}$ iterations of the Koch snowflake, for a trapping (localized) mode: $H = 0$.}
\label{fig15}
\end{figure}

Figs. \ref{fig14} and \ref{fig15} present the distributions of the following non-dimensional time-averaged energy densities: (a) kinetic $\bar{T} = \langle T \rangle/\mu $, (b) strain $\bar{W} = \langle W \rangle/\mu$, and (c) total mechanical  $\bar{E} = \bar{T} + \bar{W}$. These results correspond to propagating modes (Fig. \ref{fig14}) and trapping modes (Fig. \ref{fig15}) for the second iteration of the Koch snowflake.

As shown in Fig.~\ref{fig15}, the energy densities exhibit strong localization within the pinned region, even more pronounced than in the displacement field. This is due to their dependence on the squares of the non-dimensional displacement \eqref{Disp.Scater} and its first and second spatial derivatives \eqref{Green first derivative}. Moreover, the time-averaged energy densities are markedly higher for trapping modes than for propagating ones. This is a direct consequence of the sharper spatial variations of the displacement in the trapping regime, particularly within the pinned domain, which lead to amplified gradients and, consequently, enhanced energy localization.
   
   
\section{Concluding Remarks}
\label{sec6}		
\noindent 
This article investigates the scattering of antiplane SH waves by rigid pins embedded in an infinite two-dimensional medium, analysed within the Toupin-Mindlin strain gradient elasticity framework. Two pin configurations are studied: a fractal-based Koch snowflake and a circular arrangement with an equal number of pins along the perimeter, enabling a direct comparison between complex fractal geometry and smooth, simple shapes.

The system's resonant behavior is characterized by sharp local minima in the determinant of the Green's matrix versus the non-dimensional wavenumber, alongside a peak in its second derivative. At resonance, wave energy and particle motion concentrate predominantly within the pinned region  leading to localised trapping modes. Importantly, the response depends primarily on the geometry and boundary features rather than the sheer number of pins. Differences in resonance between the Koch snowflake and circular configurations stem from their distinct geometric and physical properties. The Koch snowflake's fractal boundary with its persistent sharp corners and cusps facilitates multiple internal reflections, promoting strong localized wave trapping and sharper resonance peaks that intensify with each iteration. In contrast, the smooth, convex circular boundary disperses waves more evenly significantly reducing localized resonance and yielding broader, less distinct peaks in the determinant of the Green's matrix.

Additionally, the ratio of the microstructural characteristic lengths crucially affects dispersion, influencing both the qualitative emergence and quantitative intensity of resonances. Trapping becomes more pronounced in the anomalous dispersion regime. High micro-inertia shifts the system toward normal dispersion, acting like a shield that stiffens the medium and suppresses resonant motion amplitudes.

These findings align with previous studies on wave interaction with rigid pins in Kirchhoff plate theory and have potential applications in wave trapping technologies such as energy harvesting, vibration isolation, and the design of acoustic metamaterials for tailored wave control.

\begin{appendices}
\numberwithin{figure}{section}
\numberwithin{equation}{section}
\section[Derivation of the Green's function]{Derivation of the Green's function}
\label{A}

\noindent
The detailed derivation of the Green's function presented in Eq. \eqref{Green} is provided below. The solution to the differential equation \eqref{THPDE} will be obtained using the double Fourier transform, which is defined as
\begin{equation}\label{eq:A.1}
\bar{f}(\xi,\eta) = \int_{-\infty}^{+\infty} \int_{-\infty}^{+\infty} f(x,y) e^{i (\xi x + \eta y)} \,dx\,dy.
\end{equation}
The inverse double Fourier transform is respectively defined as
\begin{equation}\label{eq:A.2}
f(x,y) = \frac{1}{4 \pi^2} \int_{-\infty}^{+\infty} \int_{-\infty}^{+\infty} \bar{f}(\xi,\eta) e^{-i (\xi x + \eta y)} \,d\xi\,d\eta.
\end{equation}
For simplicity, and without loss of generality, we assume that the concentrated unit body force is applied at the origin (i.e., $\mathbf{r}' = \mathbf{0}$) so that $F_{z}=-\delta(x)\delta(y)$. Afterwards, a coordinate system shift will be performed, allowing the Green's function to be extended to the general case where the body force is applied at any arbitrary point. Applying the double Fourier transform to Eq. \eqref{THPDE} yields the following expression for the transformed displacement
\begin{equation}\label{eq:A.3}
\bar{w} = \frac{1}{\mu \ell^2(q_1^2+q_2^2)} \left(\frac{1}{k^2-q_1^2}-\frac{1}{k^2+q_2^2} \right) \, ,
\end{equation}
where $k^2=\xi^2+\eta^2$ and $(q_1,q_2)$ are given in Eq. \eqref{eq.28}-\eqref{eq.29}. By applying the inverse double Fourier transform to Eq. \eqref{eq:A.3}, we obtain
\begin{equation}\label{eq:A.4}
w= \frac{1}{4 \mu \pi^2 \ell^2} \int_{-\infty}^{+\infty} \int_{-\infty}^{+\infty} \frac{1}{q_1^2+q_2^2} \left(\frac{1}{k^2-q_1^2}-\frac{1}{k^2+q_2^2} \right)\,e^{-i (\xi x + \eta y)} \,d\xi\,d\eta.
\end{equation}
To facilitate the evaluation of the integral in Eq. \eqref{eq:A.4}, it is convenient to convert the spatial coordinates $(x,y)$ and the wavenumber components $(\xi,\eta)$ into polar coordinates. The transformations are given by:
\begin{equation}\label{eq:A.5}
x = r \cos(\theta), \quad y = r \sin(\theta), 
\end{equation}
\begin{equation}\label{eq:A.6}
\xi = k \cos(\phi), \quad \eta = k \sin(\phi),
\end{equation}
and the integral in \eqref{eq:A.4} can be written as
\begin{equation}\label{eq:A.7}
w = \frac{1}{4 \mu \pi^2 \ell^2}  \int_0^{2 \pi} \int_0^{\infty} \frac{k}{q_1^2+q_2^2} \left(\frac{1}{k^2-q_1^2}-\frac{1}{k^2+q_2^2} \right) \, e^{-i k r \cos(\phi-\theta)} \,dk\,d\phi,
\end{equation}
with $r^2=x^2+y^2$.
The last expression can be simplified further by noting that
\begin{equation}\label{eq:A.8}
J_0(kr) = \frac{1}{2 \pi} \int_0^{2 \pi} e^{-i k r \cos(\phi - \theta)} \,d\phi,
\end{equation}
where $J_0(kr)$ is the Bessel function of the first kind of order zero. In particular, we have then
\begin{equation}\label{eq:A.9}
w(r) = \frac{1}{2 \mu \pi \ell^2}  \int_0^{\infty} \frac{k}{q_1^2+q_2^2} \left(\frac{1}{k^2-q_1^2}-\frac{1}{k^2+q_2^2} \right) \, J_0(kr) \,dk.
\end{equation}
The integrals in Eq. \eqref{eq:A.9} can be computed using the well-known identities \cite{watson1922treatise}
\begin{equation}\label{eq:A.10}
\int_0^\infty \frac{k}{k^2 - q_1^2} J_0(kr) \,dk = \frac{i \pi}{2} H_0^{(1)}(q_1 r) , \qquad 
\int_0^\infty \frac{k}{k^2 + q_2^2} J_0(kr) \,dk = K_0(q_2 r) \, ,
\end{equation}
where $H_0(r)$ is the first kind Hankel function of zeroth order and $K_0(r)$ is the second kind modified Bessel function of zeroth order. Note that $H_0^{(1)}$ is properly combined with the time-harmonic response $e^{-i \omega t}$ to produce outgoing propagating waves for $r>0$ whereas $K_0(r)$ describes the exponentially decaying evanescent modes. The Green's function given in Eq. \eqref{Green} is finally obtained by substituting Eqs. \eqref{eq:A.10} into Eq. \eqref{eq:A.9}.
\section[Properties of the Koch Snowflake]{Properties of the Koch Snowflake}
\label{B}

\noindent
The Koch snowflake exhibits several important geometric properties, which are outlined below:

In each iteration the number of sides and corners is increased fourfold compared to the previous iteration, thus after $n$ iterations the number of sides and corners is given by
\begin{equation}\label{eq:B.1}
N_{n} = 3 \cdot 4^n, \quad n \geq 0.
\end{equation}
If the original equilateral triangle has sides of length $S_0$, then the length of each side of the snowflake after $n$ iterations is
\begin{equation}\label{eq:B.2}
S_{n} = \frac{S_{n - 1}}{3} = \frac{S_0}{3^n}, \quad n \geq 1.
\end{equation}
Since all sides of the snowflake are of equal length, the perimeter after $n$ iterations is obtained by multiplying the number of sides $N_n$ by the length of each side $S_n$, i.e.
\begin{equation}\label{eq:B.3}
P_{n} = N_{n} \cdot S_n = 3 \cdot S_0 \cdot \left( \frac{4}{3} \right)^n, \quad n \geq 0,
\end{equation}
while the total area of the snowflake after $n$ iterations is given by
\begin{equation}\label{eq:B.4}
A_n = \frac{a_{0}}{5} \left[ 8 - 3 \left(\frac{4}{9} \right)^n \right], \quad n \geq 1,
\end{equation}
where $a_{0}$ is the area of the equilateral triangle.

From Eqs. \eqref{eq:B.3}, \eqref{eq:B.4} it is clear that as $n \rightarrow \infty$, the perimeter of the snowflake becomes unbounded, while the enclosed area remains finite.

The fractal dimension of the Koch snowflake can be characterized using the Minkowski–Bouligand (box-counting) dimension. For a subset \( S \subset \mathbb{R}^2 \), this is defined as
\begin{equation}\label{eq:B.5}
\dim_{\text{box}}(S) := \lim_{\varepsilon \to 0} \left[ \frac{\ln N(\varepsilon)}{\ln(1/\varepsilon)} \right],
\end{equation}
where \( N(\varepsilon) \) is the number of boxes of side length \( \varepsilon \) required to cover the set \( S \). For the Koch snowflake after \( n \) iterations, the number of segments along the boundary is \( N(\varepsilon_n) = 3 \cdot 4^n \), and the length of each segment is \( \varepsilon_n = S_0 / 3^n \).

Setting $S_0 = 1$ and substituting into Eq.~\eqref{eq:B.5}, we obtain an effective box-counting dimension at finite iteration \( n \):
\begin{equation}\label{eq:B.6}
\dim_{\text{box}}^{(n)} = \frac{\ln(3 \cdot 4^n)}{\ln(3^n)} = \frac{1}{n} + \frac{\ln 4}{\ln 3}.
\end{equation}
This expression shows that the box-counting dimension decreases monotonically with increasing \( n \) and converges to the classical fractal dimension of the Koch snowflake as \( n \to \infty \),
\begin{equation}
\dim_{\text{box}} = \lim_{n \to \infty} \dim_{\text{box}}^{(n)} = \frac{\ln 4}{\ln 3} \approx 1.2619.
\end{equation}
Thus, even at finite iterations, the snowflake exhibits a non-integer dimension, capturing the scale complexity of its boundary.

\end{appendices}

\section*{Acknowledgments}
\addcontentsline{toc}{section}{Acknowledgments}
E. Alevras gratefully acknowledges the financial support of ELKE at NTUA through the award of a scholarship.

\bibliographystyle{elsarticle-num}
\bibliography{References.bib}

\end{document}